\documentclass[journal,10pt]{IEEEtran}

\usepackage{graphicx}
\usepackage{enumerate}
\usepackage{cite}
\usepackage{bm}
\usepackage{color}
\usepackage{amsmath,amssymb,amsthm}
\usepackage{setspace}

\usepackage[linesnumbered,boxruled,vlined]{algorithm2e}
\usepackage[hidelinks]{hyperref}
\usepackage{flushend}

\makeatletter
\renewcommand{\Indentp}[1]{%
  \advance\leftskip by #1
  \advance\skiptext by -#1
  \advance\skiprule by #1}%
\renewcommand{\Indp}{\algocf@adjustskipindent\Indentp{\algoskipindent}}
\renewcommand{\Indpp}{\Indentp{0.5em}}%
\renewcommand{\Indm}{\algocf@adjustskipindent\Indentp{-\algoskipindent}}
\renewcommand{\Indmm}{\Indentp{-0.5em}}%
\let\oldnl\nl
\newcommand{\nonl}{\renewcommand{\nl}{\let\nl\oldnl}}
\makeatother

\usepackage[font=footnotesize,skip=6.4pt]{caption}
\usepackage{config}

\begin{document}

\title{Beyond Diagonal RIS for Multi-Band Multi-Cell MIMO Networks: A Practical Frequency-Dependent Model and Performance Analysis}
\author{Arthur S. de Sena, \textit{Member}, \textit{IEEE}, Mehdi Rasti, \textit{Senior Member}, \textit{IEEE}, Nurul H. Mahmood,\\Matti Latva-aho, \textit{Fellow}, \textit{IEEE}

\thanks{Arthur S. de Sena, Mehdi Rasti, Nurul H. Mahmood, and Matti Latva-aho are with the University of Oulu, Oulu, Finland (email: arthur.sena@oulu.fi, mehdi.rasti@oulu.fi, nurulhuda.mahmood@oulu.fi, matti.latva-aho@oulu.fi).}
\thanks{The research leading to this paper received support from the Smart Networks and Services Joint Undertaking (SNS JU) under the European Union’s Horizon Europe research and innovation programme within \href{https://hexa-x-ii.eu/}{Hexa-X-II project} (Grant Agreement No 101095759), the Academy of Finland under the \href{https://www.6gflagship.com/}{6G Flagship program} (Grant No 346208) and the academy project ReWIN-6G (Grant No 357120).}
}


\maketitle

\begin{abstract}
This paper delves into the unexplored frequency-dependent characteristics of beyond diagonal reconfigurable intelligent surfaces (BD-RISs). 
A generalized practical frequency-dependent reflection model is proposed as a fundamental framework for configuring fully-connected and group-connected RISs in a multi-band multi-base station (BS) multiple-input multiple-output (MIMO) network.
Leveraging this practical model, multi-objective optimization strategies are formulated to maximize the received power at multiple users connected to different BSs, each operating under a distinct carrier frequency.
By relying on matrix theory and exploiting the symmetric structure of the reflection matrices inherent to BD-RISs, relaxed tractable versions of the challenging problems are achieved for scenarios with obstructed and unobstructed direct channel links. The relaxed solutions are then combined with codebook-based approaches to configure the practical capacitance values for the BD-RISs. 
Simulation results reveal the frequency-dependent behaviors of different RIS architectures and demonstrate the effectiveness of the proposed schemes. Notably, BD-RISs exhibit high reflection performance across the intended frequency range, remarkably outperforming conventional single-connected RISs.
Moreover, the proposed optimization approaches prove effective in enabling the targeted operation of BD-RISs across one or more carrier frequencies.
The results also shed light on the potential for harmful interference in the absence of synchronization between RISs and adjacent BSs.

\end{abstract}



\section{Introduction}

The run toward the sixth generation (6G) of wireless communication has already started. The relentless quest for seamless coverage, ever-increasing data rates, and high spectral and energy efficiencies is driving the research and development of novel and transformative technologies. In the vanguard of 6G, \ac{RIS} has emerged as a revolutionary paradigm with the promising capability of optimizing the propagation environment, a component over which conventional wireless systems have little or no control. Engineered with a large number of nearly passive reflecting elements with software-tunable electromagnetic properties, an \ac{RIS} can dynamically adapt and manipulate propagation channels to realize diverse functions, ranging from signal coverage extension and rate maximization to enhanced control of user channel gains and interference mitigation in multi-user scenarios \cite{Sena20, Sena22}.

A conventional \ac{RIS} can be modeled as a reconfigurable impedance network, where each reflecting element is connected to a single impedance (a reconfigurable self-impedance), which is independent of the other elements. Due to this characteristic, conventional \acp{RIS} are also called single-connected \acp{RIS}. One implication of this single-connected \ac{RIS} architecture is that the induced reflections are traditionally modeled by a diagonal reflection matrix.
Recently, as an attempt to improve the reflection efficiency of conventional \acp{RIS}, the authors of \cite{Shen22} proposed a generalized architecture inspired by multi-port network theory. In this new architecture, each \ac{RIS} reflecting element, in addition to being connected to its self-impedance, can be also interconnected with the other elements through internal impedances, giving rise to two new \ac{BD-RIS} concepts: fully-connected \ac{RIS} and group-connected \ac{RIS}. In a fully-connected \ac{RIS}, all reflecting elements are interconnected and their induced reflections are now modeled by a full matrix, whereas in a group-connected \ac{RIS} only sub-groups of the elements are fully-connected but independent across groups, thus, resulting in a block-diagonal reflection matrix. The fully-connected \ac{RIS} offers the highest reflected power but also counts with the highest circuit complexity. On the other hand, the group-connected \ac{RIS} provides an intermediate reflection performance, ranging between the performance levels of the conventional single-connected \ac{RIS} and the fully-connected \ac{RIS}, but with a lower overall complexity.

In contrast to legacy active radio technologies, both single-connected and \ac{BD-RIS} architectures require only low-power low-cost components to enable the reconfigurability of their reflecting elements. This attractive characteristic positions \acp{RIS} as plug-and-play devices with the potential for easy deployment across diverse environments in the future 6G landscape. However, this versatility, while promising, also raises significant concerns.
For instance, as we move towards the 6G era, the trend of network densification, characterized by the increasing proximity of \acp{BS}, is expected to intensify. Additionally, the rising popularity of local networks, combined with the growing necessity of multi-frequency operation, driven by the demanding spectrum requirements forecasted for 6G, should increase the likelihood of multiple \acp{BS} under multiple bands, controlled by either a single or multiple operators, coexisting within the same spatial location. Assisting such scenarios with \acp{RIS} may introduce as well serious problems that deserve attention. Specifically: 
\begin{itemize}\setlength\itemsep{1mm}
    \item Modeling and optimizing \acp{RIS} without considering their frequency dependency characteristics can result in unexpected effects in practical multi-band environments. This issue may hinder the full potential of \ac{RIS} technology and lead to degraded communication performance in real-world systems.
    
    \item The development of practical configuration strategies for enabling the multi-frequency operation of emerging \ac{BD-RIS} architectures is another urgent and critical challenge.
    The highly coupled reflecting elements of BD-RISs make their tuning inherently more complex than existing solutions for conventional single-connected RISs. Consequently, while BD-RISs offer enhanced reflection efficiency, they also bring a set of new challenges that must be addressed to effectively serve users operating under different frequencies in upcoming 6G scenarios.
    %
    %

    \item Moreover, it is often assumed that the RIS optimization and channel estimation at multiple BSs happen synchronously, preventing the RIS from introducing interference or impacting the accuracy of acquired channel estimates. In practice, however, this synchronization is not always achieved. If the RIS is not synchronized or does not communicate with a neighboring BS, a likely situation in multi-operator scenarios, active channel aging may be introduced, as highlighted in \cite{Huang23, Sena24} from a physical layer security perspective. Thus, the unplanned deployment of RISs in multi-BS networks can reduce the effectiveness of channel estimation and introduce harmful interference against users in non-intended systems.

\end{itemize}

Although efforts have been made to optimize the performance of conventional single-connected RISs in multi-band environments \cite{Cai20, Li21, Cai21, Jiang22, Yashvanth23, Inwood23, Cai22}, as of now, the potential unintended out-of-band degradation impacts in multi-BS scenarios and the practical configuration of BD-RISs remain unexplored.

\subsection{Related Works}
A few papers have investigated the practical configuration and frequency response characteristics of single-connected \acp{RIS}. The work in \cite{Cai20}, for instance, studied the phase-amplitude-frequency relationship of reflected signals and introduced a practical reflection model for single-connected \acp{RIS} considering a point-to-point single-antenna \ac{OFDM} system. The same frequency-dependent model was later extended to \ac{OFDM}-aided \ac{MIMO} systems in \cite{Li21}. In both works, the authors evaluated the efficacy of the proposed model with its incorporation into the \ac{RIS} optimization. Simulation results demonstrated that significant performance improvements can be achieved by employing this practical model when compared to the cases with a frequency-blind model in multi-band scenarios.
The frequency-dependency of single-connected \acp{RIS} in multi-band multi-cell networks was addressed in \cite{Cai21} and \cite{Cai22}. By assuming that each \ac{BS} operates under a different frequency and adopting a simplified frequency-dependent reflection model, the authors jointly designed the \ac{BS} precoders and \ac{RIS} phase-shift matrices aiming at minimizing the total transmit power.
Heterogeneous multi-band networks assisted by practical single-connected \acp{RIS} were studied in \cite{Jiang22}, where a frequency-aware iterative algorithm was developed to maximize the sum rate of all users under each operating frequency.
In \cite{Yashvanth23}, the out-of-band performance of a multi-user single-antenna \ac{RIS}-assisted system was investigated considering a frequency-dependent analytical reflection model, 
and the work in \cite{Inwood23} optimized a single-connected RIS structured into sub-surfaces to assist a multi-frequency \ac{MIMO} system, where each sub-surface was assigned to a distinct frequency.

Recent studies focusing on diverse aspects of BD-RISs also exist. However, none has investigated their practical frequency-dependent behavior and configuration.
For example, the pioneering work on \ac{BD-RIS} in \cite{Shen22}, investigated the scaling law of the received signal power of both fully-connected and group-connect \acp{RIS}, and showed that the two architectures can significantly improve the power reflection efficiency, remarkably outperforming conventional single-connected RISs. Nevertheless, the influence of frequency was not addressed. 
In \cite{Li23}, the authors introduced a novel concept of group-connected BD-RIS with a dynamic grouping strategy, in which the reflecting elements were adaptively divided into subsets based on the observed \ac{CSI} of users within a \ac{MIMO} system. Simulation results demonstrated the performance superiority of the proposed dynamic scheme over static approaches.
In \cite{LiMS23} and \cite{LiTR23}, the architecture capabilities of \acp{BD-RIS} were generalized from only reflective to transmissive and hybrid modes, where double-sided or multi-sector \acp{BD-RIS} can operate reflecting, transmitting, or executing both functions simultaneously.
The work in \cite{Li22} proposed a new \acp{BD-RIS} design, in which the reflecting elements were interconnected via switch arrays. The architecture exhibited a lower optimization complexity than that imposed by fully and group-connected RISs. Moreover, in the presented results, the proposed switch-based RIS outperformed the group-connected RIS and approached the performance of the fully-connected RIS counterpart.
The combination of \ac{BD-RIS} with \ac{RSMA} was considered in \cite{Soleymani23} and \cite{Fang22}. The authors of \cite{Soleymani23}, specifically, developed a general optimization framework to enhance the spectral and energy efficiencies of both fully and group-connected \ac{RIS}-aided \ac{RSMA} in a \ac{URLLC} system, while the work in \cite{Fang22} concentrated on the application of only fully-connected \acp{RIS} to the downlink of a multi-antenna \ac{RSMA} scheme. 
Low-complexity closed-form optimization strategies for \acp{BD-RIS} were proposed in \cite{Fang23} and \cite{NeriniCF23},
and the impact of discrete coefficients on the performance of \acp{BD-RIS} was studied in \cite{Nerini23}. 
In \cite{Santamaria23}, the \ac{SNR} maximization problem was explored for single-antenna and MIMO channels assisted by \acp{BD-RIS}, and \cite{NeriniPF23} derived the Pareto frontier for the performance-complexity trade-off of different \ac{BD-RIS} architectures.

\subsection{Motivation and Contributions}\vspace{-2mm}
As described in the previous subsection, the performance advantages of \acp{BD-RIS} have been demonstrated across diverse scenarios and applications. Nevertheless, while a few works on single-connected \ac{RIS} have taken multi-band frequencies into consideration, existing research on \ac{BD-RIS} exclusively focuses on ideal frequency-blind reflection models.
Addressing this lacuna, this paper extends the results of \cite{Shen22} by investigating for the first time the frequency-dependent behavior of \acp{BD-RIS}. Further details and the main contributions of this work are summarized as follows:\vspace{-2mm}

\begin{itemize}\setlength\itemsep{1mm}
    \item Building upon the contributions of \cite{Cai21, Jiang22}, which focused only on single-connected RISs, we propose a novel and generalized practical frequency-dependent reflection model applicable for \acp{BD-RIS}, i.e., fully-connected and group-connected RISs, which is fine-tuned to exhibit optimized performance for frequencies between $4$~GHz and $12$~GHz. Frequencies within this range are currently under active investigation by the industry and have been identified by the International Telecommunication Union (ITU) as future candidates for 6G services due to their favorable propagation characteristics \cite{cui20236g}. This practical circuit model is used as our basic framework for configuring the deployed \ac{BD-RIS} in a multi-band multi-\ac{BS} \ac{MIMO} environment.
    
    \item We formulate multi-objective strategies to optimize the practical reflecting coefficients of both fully-connected and group-connected RISs for assisting multiple users connected to different multi-antenna \acp{BS}, with each \ac{BS} operating under a distinct carrier frequency. The formulated problems are difficult to solve directly due to the frequency-dependent behavior and the highly coupled reflection coefficients inherent to BD-RISs. To tackle the challenging original formulation, we rely on matrix theory and exploit the symmetric structure of the reflection matrix of \acp{BD-RIS} to achieve relaxed tractable versions of the problems.
    Aiming at achieving fundamental insights into the performance of BD-RISs, we start by considering the scenario where direct channel links between the BS and users are under deep fading, e.g., due to severe blockage. In this first case, by applying a series of matrix transformations, we obtain convex equivalent relaxed problems for fully-connected and group-connected RISs that can be solved in closed form with the aid of the singular value decomposition (SVD).
    
    \item Next, we extend our relaxed optimization strategies to scenarios where the signal contributions in the direct links are non-negligible. In these more complex scenarios, closed-form solutions are not available. Alternatively, we propose efficient conditional gradient-based algorithms to solve the generalized problems. The relaxed solutions for the scenarios with both blocked and available direct links are mapped to practical capacitance values to configure the BD-RISs. This mapping is facilitated by impedance codebooks constructed based on the proposed frequency-dependent circuit models, relying on microwave theory.

    \item Comprehensive simulation results are provided to investigate the frequency-dependent behavior of different \ac{RIS} architectures and to validate the effectiveness of the developed practical optimization strategies. For instance, our results show that the proposed \acp{BD-RIS} circuit model exhibits a high performance when operating under intended frequencies, which can remarkably outperform conventional single-connected \ac{RIS} counterparts across almost the entire considered frequency range. Furthermore, it is demonstrated that the proposed optimization strategies are effective and enable the RISs to target one or more operating frequencies. Our results also show that harmful interference can be generated for adjacent systems if their BSs, i.e., their channel estimation processes, are not synchronized with the \ac{RIS} configuration.
\end{itemize}

\vspace{1mm}

\noindent  {\it Notation:} Bold-faced lower-case letters denote vectors and upper-case represent matrices. The $i$th element of a vector $\mathbf{a}$ is denoted by $[\mathbf{a}]_i$, the $(ij)$ entry of a matrix $\mathbf{A}$ by $[\mathbf{A}]_{ij}$, the submatrix of $\mathbf{A}$ formed by its rows (columns) from $i$ to $j$ by $[\mathbf{A}]_{i:j,:}$ $\left([\mathbf{A}]_{:,i:j}\right)$. The $L_2$ norm of a vector $\mathbf{a}$ is denoted by $\|\mathbf{a}\|_2$, and the Frobenius norm of a matrix $\mathbf{A}$ by $\|\mathbf{A}\|_F$. The transpose, Hermitian transpose, and inverse of $\mathbf{A}$ are represented by $\mathbf{A}^T$, $\mathbf{A}^H$, and $\mathbf{A}^{-1}$, respectively, $\mathbf{I}_M$ is the $M\times M$ identity matrix, $\mathbf{0}_{M, N}$ is the $M\times N$ zero matrix, and $\otimes$ represents the Kronecker product. The operator $\mathrm{vec}(\cdot)$ stacks the columns of an $M\times N$ matrix into a column vector of length $MN$, $\mathrm{unvec}(\cdot)$ applies the inverse operation, $\mathrm{vech}(\cdot)$ transforms the lower triangular half of an $M\times M$ matrix into a column vector of length $M(M+1)/2$, $\mathrm{bdiag}(\cdot)$ creates a $KM\times KM$ block diagonal matrix from $K$ passed $M \times M$ matrices, and $\mathrm{E}(\cdot)$ denotes expectation.\vspace{-3mm}

\section{System Model}
\begin{figure}[t]
	\centering
	\includegraphics[width=.98\linewidth]{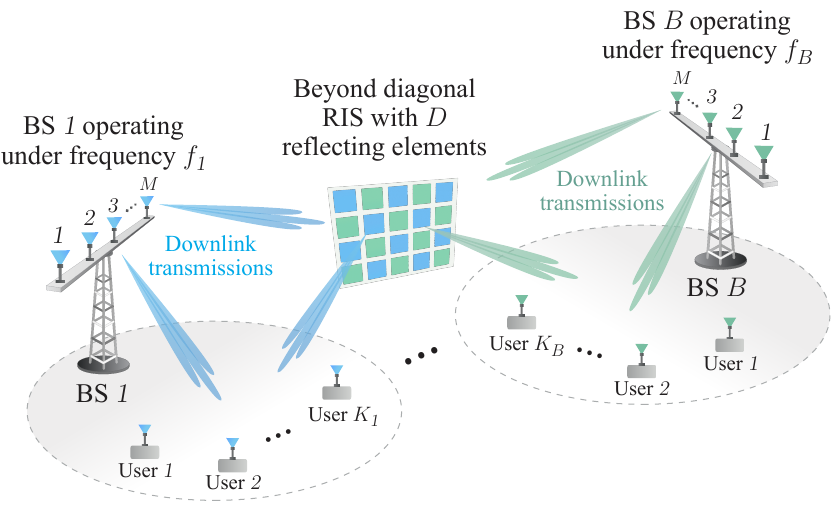}
	\caption{{System model. A multi-band multi-\ac{BS} \ac{MIMO} network is assisted by a \ac{BD-RIS}.}}\label{fig:sysmodel}
\end{figure}

We study in this work a multi-\ac{BS} downlink \ac{MIMO} network{\footnote{{Cell-free MIMO is an emerging concept with the potential to provide seamless coverage in 6G. Future work shall investigate frequency-dependent BD-RISs integrated into these promising network architectures \cite{Ma23}.}}} illustrated in Fig. \ref{fig:sysmodel}, comprising $B$ \acp{BS}, which are represented by the set $\mathcal{B} \triangleq \{1, \cdots, B\}$. We assume that the $b$th \ac{BS}, for $b=1, \cdots, B$, operates under a distinct carrier frequency $f_b$, where  $f_b \in \mathcal{F} \triangleq \{f_1, \cdots, f_B\}$, and that each \ac{BS} is equipped with $M$ transmit antennas. Moreover, there are $K_b$ single-antenna users connected to the $b$th \ac{BS}, which, in their turn, are organized in the set $\mathcal{K}_b \triangleq \{1, \cdots, K_b\}$.
Under this multi-band scenario, one \ac{BD-RIS}, comprising $D$ reflecting elements, is deployed to assist the users within the network{, where both fully-connected and group-connected RIS architectures are considered in our investigations}. To this end, practical frequency-dependent reflection models are adopted throughout this work, which will be explained in detail in Section \ref{SecBDRIS}.

Given that the BSs operate under different frequencies, we assume that users can eliminate the inter-\ac{BS} interference by employing a proper filtering strategy. This assumption enables the $b$th \ac{BS} to multiplex users in space through linear precoding based only on the channels observed under frequency $f_b$. 
To this end, we adopt in this work a narrow-band\footnote{The study of frequency-dependent BD-RIS architectures on wide-band systems arises as a potential future direction.} block-fading channel model, in which the channels remain constant within a given time coherent interval but vary independently across distinct intervals. As a result, the signal received by the $k$th user connected to the $b$th \ac{BS}, $\forall k \in \mathcal{K}_b$, $\forall b \in \mathcal{B}$, propagating through both the direct and reflected \ac{BS}-RIS-user links, can be written as
\begin{align}
    y_{bk} = \sum_{\forall u\in \mathcal{K}_b} [\mathbf{f}_{bk}^H
      \bm{\Theta}(\mathbf{C}, f_b)\mathbf{G}_{b} + \mathbf{h}_{bk}^H]\mathbf{p}_{bu} \sqrt{P \alpha_{bu} } x_{bu} + n_{bk},
\end{align}
where $x_{bu}$ represents the data symbol intended for the $u$th user, $\alpha_{bu} \in [0,1]$ is the corresponding power allocation coefficient, $P$ is the total transmit power budget, and $n_{bk}$ is the additive noise experienced at the user device, which follows the Complex Gaussian distribution with zero mean and variance $\sigma^2$. In addition, $\mathbf{p}_{bu}$ is the precoding vector, which is constructed as a zero-forcing precoder\footnote{The design of more sophisticated precoding strategies for multi-band BD-RIS-assisted environments is beyond the scope of this paper and is left for future research, where a thorough investigation can be carried out.}, such that 
\begin{align}
 [\mathbf{f}_{bk}^H
      \bm{\Theta}(\mathbf{C}, f_b)\mathbf{G}_{b} + \mathbf{h}_{bk}^H] \mathbf{p}_{bu} &= 0, & \forall u\neq k,
\end{align}
satisfying $\|\mathbf{p}_{bu}\|_2^2 = 1$. The matrix $\bm{\Theta}(\mathbf{C}, f_b) \in \mathbb{C}^{D \times D}$ is the base-band \ac{RIS} scattering matrix, which is a function of both the operating frequency $f_b \in \mathcal{F}$ and the matrix of capacitances $\mathbf{C} \in \mathbb{R}^{D\times D}$ employed in the practical circuit of the RIS, as more details will be provided in the next section. Last, $\mathbf{h}_{bk} \in \mathbb{C}^{M \times 1}$, $\mathbf{f}_{bk} \in \mathbb{C}^{D \times 1}$ and $\mathbf{G}_{b} \in \mathbb{C}^{D \times M}$ comprise the complex baseband channel coefficients for the links between the BS and the $k$th user, the \ac{RIS} and the $k$th user, and the $b$th \ac{BS} and the \ac{RIS}, respectively, $\forall u \in \mathcal{K}_b$ and $\forall b \in \mathcal{B}$, in which Rayleigh fading is considered in all channel links, modeling rich scattering environments.

\section{Frequency-Dependent Model for BD-RISs}\label{SecBDRIS}
In this section, the multi-port-based reflection model for \acp{BD-RIS} proposed in \cite{Shen22} is generalized for multi-band operation. In particular, we start by explaining the concept of a fully-connected \ac{RIS}, in which all reflecting elements are interconnected through reconfigurable impedances and, subsequently, the practical model is extended to group-connected \ac{RIS}, as follows.\vspace{-3mm}

\begin{figure}[!t]
	\centering
 \includegraphics[width=.85\linewidth]{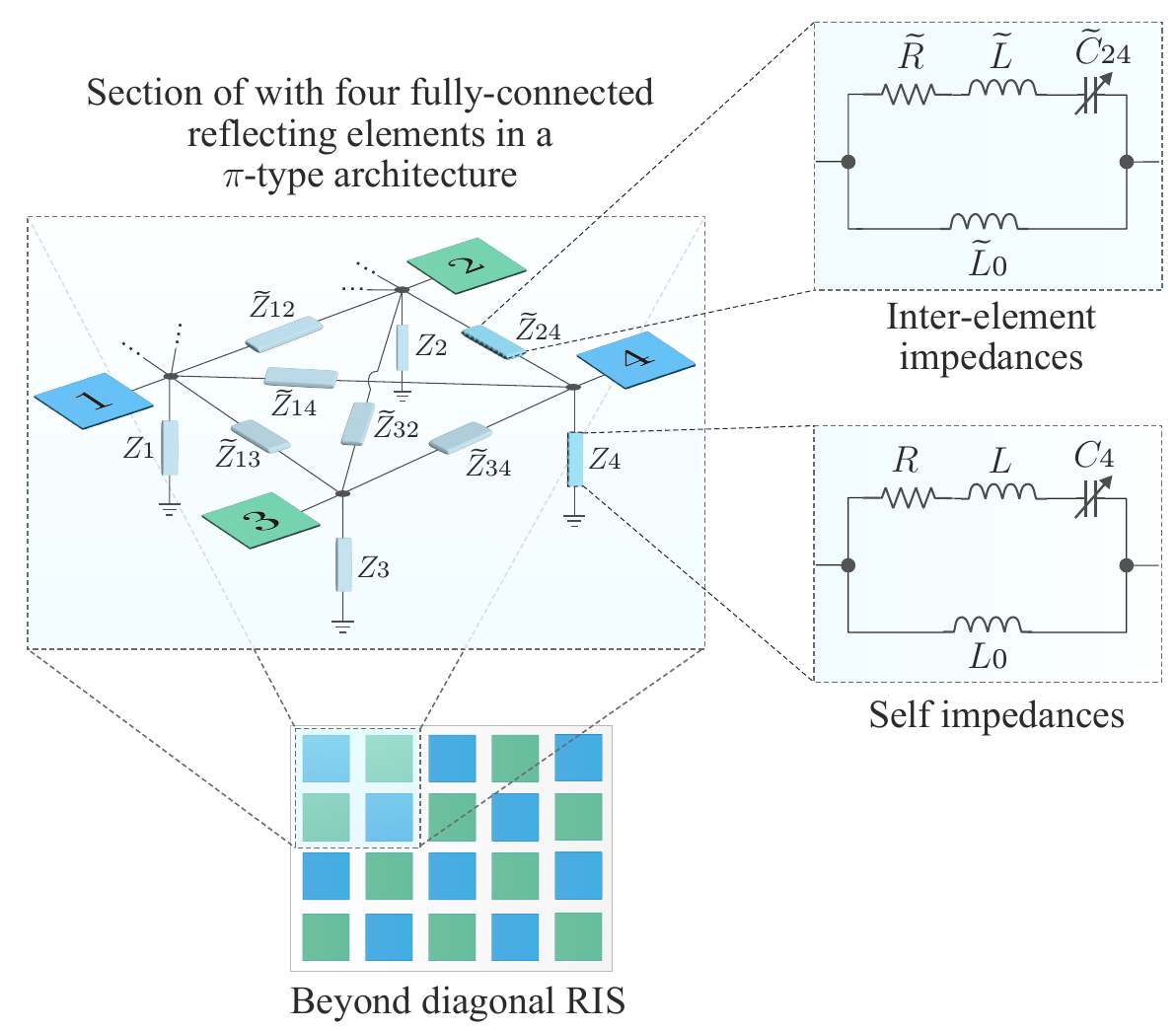}
	\caption{{Illustration of a \ac{BD-RIS} section with four elements arranged in a fully-connected $\pi$-type circuit configuration.}}\label{ris_circuit}
\end{figure}

\subsection{Fully-Connected RIS}

A fully-connected \ac{RIS} comprising $D$ reflecting elements can be treated as a $D$-port reciprocal network, where the induced reflections can be modeled by a generalized scattering matrix $\bm{\Theta} \in \mathbb{C}^{D\times D}$, which is symmetric, i.e., $\bm{\Theta} = \bm{\Theta}^T$, due to reciprocity, and satisfies $\bm{\Theta} \bm{\Theta}^H \preceq \mathbf{I}_{D}$, which implies that $\frac{1}{\sqrt{D}} \|\bm{\Theta}\|_F \leq 1$ due to conservation of energy, where the equality is achieved only when the \ac{RIS} circuits are lossless. For this ideal case, the \ac{RIS} scattering matrix becomes unitary, i.e., $\bm{\Theta} \bm{\Theta}^H = \mathbf{I}_{D}$%
. Under these observations, the \ac{RIS} scattering matrix can be given by
\begin{align}\label{scatmat1}
    \bm{\Theta} = (\mathbf{Z} + Z_0\mathbf{I}_{D})^{-1} (\mathbf{Z} - Z_0\mathbf{I}_{D}),
\end{align}
where $Z_0$ is the reference impedance of the transmission medium and $\mathbf{Z} \in \mathbb{C}^{D\times D}$ is the matrix collecting the effective impedances that relates the currents and voltages between the multiple interconnected ports of the RIS \cite{Pozar12}. Specifically, the transfer impedance matrix for a fully-connected multi-port network can be given by $\mathbf{Z} = \mathbf{Y}^{-1}$, where $\mathbf{Y}$ denotes the transfer admittance matrix of the corresponding circuit. In particular, the circuit architecture of the BD-RISs considered in this work is modeled as a fully-connected $\pi$-type equivalent network, as illustrated in Fig. \ref{ris_circuit}. For $\pi$-type networks, the entries of $\mathbf{Y}$ can be calculated as follows \cite{Nie14, Morched93}:
\begin{align}\label{admit_mat}
    [\mathbf{Y}]_{pq} = \begin{cases} -\tilde{Z}_{pq}^{-1}, & \text{if} \quad p \neq q \\[1mm] 
    Z_{p}^{-1} + \sum_{i=1, i\neq p}^D \tilde{Z}_{pi}^{-1} & \text{if} \quad p = q,
    \end{cases}
\end{align}
where $Z_{p}$ represents the self-impedance connecting the port $p$ to the ground, $\forall p \in \{1, \cdots, D\}$, and $\tilde{Z}_{pq}$ is the internal impedance connecting ports $p$ and $q$, $\forall q \neq p \in \{1, \cdots, D\}$, such that $\tilde{Z}_{pq} = \tilde{Z}_{qp}$, due to reciprocity.
This implies that the entries of the scattering matrix $\bm{\Theta}$ can be completely determined by properly tuning $\frac{D(D+1)}{2}$ reconfigurable impedances (or, equivalently, admittances).

\begin{figure*}[!t]
	\centering
 \includegraphics[width=.8\textwidth]{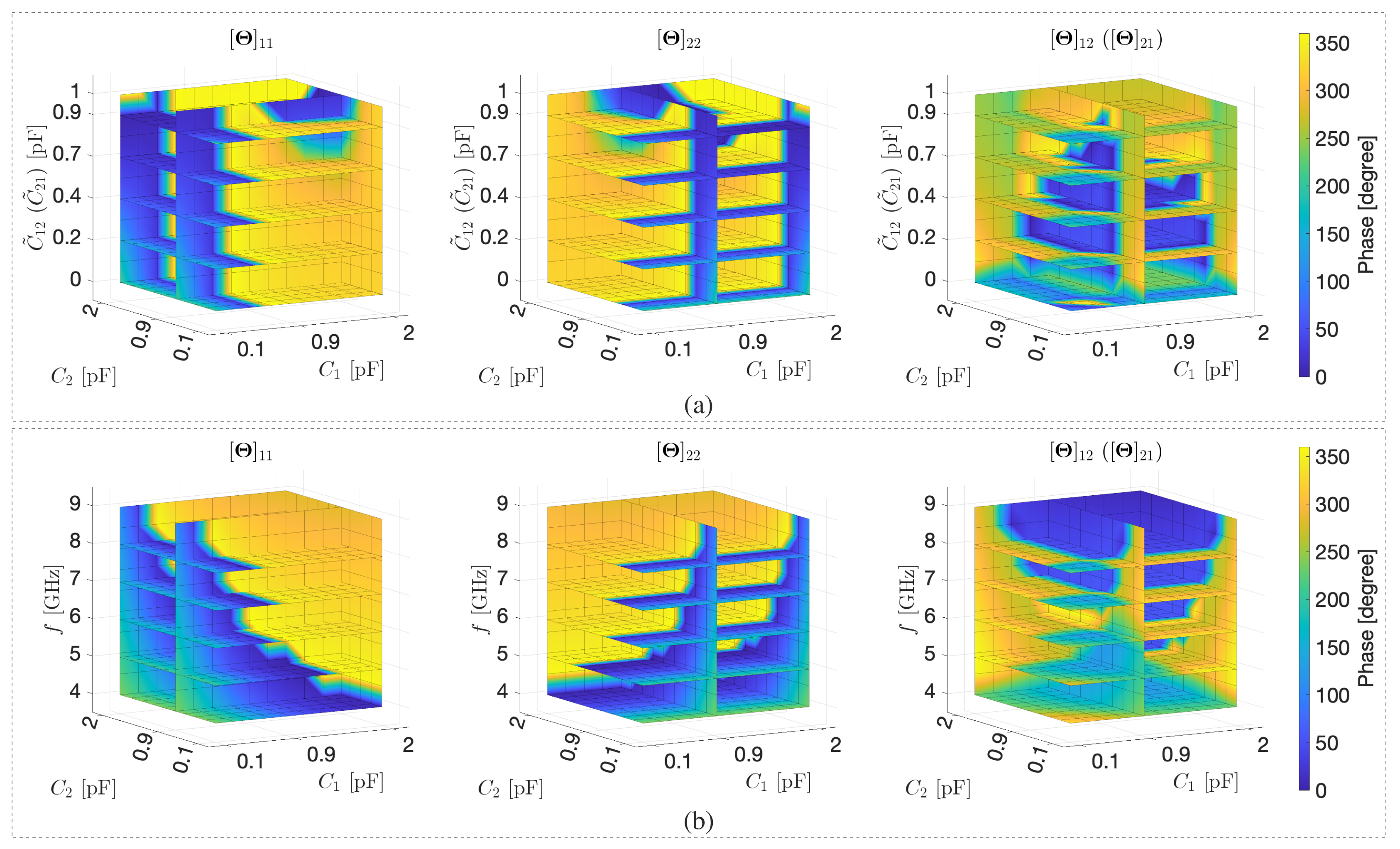}
	\caption{Four-dimensional visualization of the phases of the scattering coefficients for a two-element fully-connected RIS with the self-capacitances for ports $1$ and $2$ represented in the $x$-axis and $y$-axis, respectively, considering: (a) a frequency of $f = 7$~GHz and the inter-element capacitance in the $z$-axis (the vertical axis), and (b) a fixed inter-element capacitance of $\Tilde{C}_{12} = \Tilde{C}_{21} = 0.2$~pF with different frequency values in the $z$-axis ($R = \tilde{R} = 1$~$\Omega$, $L_0 = 2.5$~nH, $L = 0.7$~nH, $\tilde{L}_0 = 12.5$~nH, $\tilde{L} = 0.2$~nH, and $Z_0 = 50$).}\label{phase_resp}\vspace{-2mm}
\end{figure*}

{The} self-impedances for each reflecting element can be modeled as a parallel resonant circuit \cite{Cai20, Jiang22}, illustrated in Fig. \ref{ris_circuit}{. An expression for the equivalent self-impedance $Z_{p}$ for such a circuit can be easily obtained by relying on basic circuit theory. To be more specific, by recalling Kirchhoff’s current law and Ohm's law, we can readily achieve}
{\begin{align}\label{selfimp_eq_0}
    \frac{1}{Z_{p}} = \frac{1}{Z_{L_0}} + \frac{1}{Z_{L} + Z_{C_p} + R} = \frac{Z_{L_0} + Z_{L} + Z_{C_p} + R}{Z_{L_0}(Z_{L} + Z_{C_p} + R)},
\end{align}
}{where $Z_{L_0}$ and $Z_{L}$ denote the impedances of the inductors $L_0$ and $L$, respectively, corresponding to the inner and outer layers of the \ac{RIS} reflecting element, $Z_{C_p}$ is the impedance associated with the reconfigurable capacitance $C_{p}$, and $R$ is an effective resistance modeling the losses of the circuit. By replacing $Z_{L_0}$, $Z_{L}$, and $Z_{C_p}$ in \eqref{selfimp_eq_0} with the actual impedances of the corresponding circuit components, we can obtain the following expression for the $p$th self-impedance:}
\begin{align}\label{selfimp_eq}
    Z_{p}(C_{p}, f) &= \frac{j 2 \pi f L_0 \Big(j 2 \pi f L + \frac{1}{j2 \pi f C_{p}} + R\Big)}{j 2 \pi f L_0 + j 2 \pi f L + \frac{1}{j2 \pi f C_{p}} + R},
\end{align}
{where $f$ denotes the carrier frequency.}

{In their turn}, the reconfigurable internal impedances in the transmission lines connecting the elements are implemented by a similar resonant circuit\footnote{Note that our goal in this work is not to propose an optimal hardware architecture but to shed light on the frequency-dependent behavior of \acp{BD-RIS}. In-depth studies on circuit design for these new architectures are still missing in the literature and go beyond the scope of this paper.}, also illustrated in Fig. \ref{fig:sysmodel}, in which the upper components connected in series are the equivalent circuit for a varactor \cite{Cai20}. Like previously, we can exploit Kirchhoff’s law to promptly calculate the equivalent inter-element impedance between ports $p$ and $q$, resulting in the following:
\begin{align}\label{intelimp_eq}
    \tilde{Z}_{pq}(\tilde{C}_{pq}, f) &= \frac{j 2 \pi f \tilde{L}_0\left(j 2 \pi f \tilde{L} + \frac{1}{j2 \pi f \tilde{C}_{pq}} + \tilde{R}\right)}{j 2 \pi f \tilde{L}_0 + j 2 \pi f \tilde{L} + \frac{1}{j2 \pi f \tilde{C}_{pq}} + \tilde{R}},
\end{align}
where $\tilde{C}_{pq}$ is the capacitance between ports $p$ and $q$, $\tilde{L}$ and $\tilde{R}$ are{, respectively, the inductance and resistance associated with the varactor part of the circuit, and $\tilde{L}_0$ is an auxiliary fixed inductance that helps to fine-tune the impedance range in} the inner transmission lines. As can be seen, \eqref{selfimp_eq} and \eqref{intelimp_eq} are frequency-dependent and can be adjusted by properly tuning the reconfigurable capacitances $C_{p}$ and $\tilde{C}_{pq}$, respectively, which, throughout this paper, are organized into the matrix $\mathbf{C}$, such that the main diagonal entries are associated with \eqref{selfimp_eq} and the off-diagonal entries with \eqref{intelimp_eq}.

It is noteworthy that the frequency characteristics of each reflecting element of a fully-connected RIS cannot be analyzed independently, as performed in previous works such as in \cite{Cai20, Li21, Cai21, Jiang22, Yashvanth23, Inwood23, Cai22} for conventional single-connected RISs. Instead, the elements should be studied jointly due to their inherent mutual coupling. We plot in Fig. \ref{phase_resp} the phase response of a fully-connected RIS to demonstrate this coupled frequency-dependent behavior and the impact of tuning the circuit capacitances in \eqref{selfimp_eq} and \eqref{intelimp_eq}. For illustrative purposes, we consider the simplest case with two connected elements. As a result, the entries of the matrix $\bm{\Theta} \in \mathbb{C}^{2 \times 2}$ will be a function of three reconfigurable capacitances, $C_1$, $C_2$, and $\Tilde{C}_{12}$ (which is the same as $\Tilde{C}_{21}$), and the operating frequency $f$. To visualize this multi-dimensional phase response, Fig. \ref{phase_resp} presents sliced colored volumes, in which the different colors represent the observed phase coefficients, whereas the slices within the volumes correspond to different capacitance or frequency values, depending on the subfigure. Specifically, in Fig. \ref{phase_resp}(a), for a frequency of $f = 7$~GHz, we can see that almost the entire phase range can be achieved by properly tuning the capacitances. It is also evident that all three capacitances have influence over the three scattering coefficients, $[\bm{\Theta}]_{11}$, $[\bm{\Theta}]_{22}$, and $[\bm{\Theta}]_{12}$ ($[\bm{\Theta}]_{21}$), revealing the coupled behavior of fully-connected RISs. In Fig. \ref{phase_resp}(b), we fix the inter-element capacitance to $0.2$~pF and investigate the impact of the operating frequency on the phase coefficients. As can be seen, the frequency plays a significant role in the phase of all scattering coefficients. For instance, considering the coefficient $[\bm{\Theta}]_{11}$, when $C_2 = 0.1$~pF and $C_1 = 0.9$~pF, its phase observed in the frequency slice of $4$~GHz falls within the blue spectrum (around $50^\circ$). However, as the frequency exceeds $5$~GHz, the coefficient $[\bm{\Theta}]_{11}$ undergoes a rapid phase change, reaching approximately $350^\circ$, in the yellow color spectrum, considering the same capacitance values. Similar trends can be observed with the other scattering coefficients. Such a characteristic provides a strong indication that fully-connected RISs may perform sub-optimally when operating under non-designed frequency bands. This reinforces the need for a careful frequency-aware design and analysis of BD-RISs for their effective deployment.

As shown in a number of recent studies, a fully-connected \ac{RIS} can significantly outperform conventional single-connected \ac{RIS} counterparts. However, the associated high circuit complexity can be a limiting factor for its practical deployment. This motivates the consideration of alternative lower-complexity group-connected architectures, in which the reflecting elements are only partially connected. The practical frequency-dependent model presented for fully-connected RISs can be straightforwardly applied to group-connected RISs, as explained in the following subsection.

\subsection{Group-Connected RIS}
As anticipated, a tradeoff between complexity and performance can be achieved by considering group-connected \ac{RIS} architectures \cite{Shen22}, where the reflecting elements within individual groups are fully-connected but independent across groups. Consequently, the scattering matrix for a group-connected \ac{RIS} with $D$ elements, organized into $G$ groups, has the following block diagonal structure
\begin{align}
    \bm{\Theta} = \mathrm{bdiag}(\bm{\Theta}_1, \cdots, \bm{\Theta}_G) \in \mathbb{C}^{D\times D},
\end{align}
where $\bm{\Theta}_g \in \mathbb{C}^{\bar{D} \times \bar{D}}$ is the scattering matrix corresponding to $g$th independent fully-connected group, for $g = 1, \cdots, G$, with each group comprising $\bar{D} = D/G$ reflecting elements. Moreover, as for the fully-connected case, the practical frequency-dependent entries of each $\bm{\Theta}_g$ are determined through \eqref{scatmat1}, \eqref{selfimp_eq}, and \eqref{intelimp_eq}, thus, satisfying
$\bm{\Theta}_g = \bm{\Theta}_g^T$ and $\bm{\Theta}_g \bm{\Theta}_g^H \preceq \mathbf{I}_{\bar{D}}$. As a result,
the entries of $\bm{\Theta}_g$, for $g = 1, \cdots, G$, can be computed by selecting $\frac{\bar{D}(\bar{D}+1)}{2} = \frac{D}{2G} (\frac{D}{G} + 1)$ non-redundant impedances. It is noteworthy that the conventional single-connected \ac{RIS} corresponds to the special case when $G = D$, whereas the fully-connected \ac{RIS} is achieved when $G = 1$.

\section{RIS Optimization}\label{RISopt}
In this section, we present the optimization strategies for tuning the scattering matrix of multi-band \acp{BD-RIS}. Our overall objective is to maximize the received power at users subject to the proposed frequency-dependent \ac{RIS} constraints.
{However, because the entries of the matrix $\bm{\Theta}(\mathbf{C},f_b)$ will assume different values for each frequency $f_b \in \mathcal{F}$, and given that we dispose of a single \ac{RIS}, our objective cannot be fulfilled optimally in a global sense. Alternatively, we can find a set of capacitances $\mathbf{C}$ that provides a balance across the multiple frequencies and users. Such a goal for the fully-connected RIS can be expressed through the following weighted multi-objective problem{\footnote{{The incorporation of other optimization parameters into \eqref{prob_1}, like power allocation coefficients $\alpha_{bu}$, should offer greater performance gains. This interesting possibility shall be considered in a potential extension of this work.}}}:
\begin{subequations}\label{prob_1}
\begin{align}
    \underset{\mathbf{C}}{\max} \hspace{2mm} &
      \sum_{b \in \mathcal{B}} \mu_{b} \sum_{k \in \mathcal{K}_b} \nu_{k} \left\| \mathbf{f}_{bk}^H \bm{\Theta}(\mathbf{C}, f_b)\mathbf{G}_{b} + \mathbf{h}_{bk}^H
    \right\|_2^2,
    \label{prob_1a}\\[0mm]
    \text{s.t.} \hspace{3mm} & \bm{\Theta}(\mathbf{C},f_b) \bm{\Theta}(\mathbf{C},f_b)^H \preceq \mathbf{I}_{D} \label{prob_1b},\\
    & \hspace{0mm} \bm{\Theta}(\mathbf{C},f_b) = \bm{\Theta}(\mathbf{C},f_b)^T \label{prob_1c},
\end{align}
\end{subequations}
where $\mu_{b}$ and $\nu_{k}$ are the optimization weights associated with the BSs and the corresponding connected users, respectively.

As for the group-connected RIS, the following analogous problem can be formulated:
\begin{subequations}\label{prob_rev_2}
\begin{align}
    &\underset{\mathbf{C}}{\max} \hspace{2mm}
     \sum_{b \in \mathcal{B}} \mu_{b} \sum_{k \in \mathcal{K}_b} \nu_{k} \left\| \mathbf{f}_{bk}^H \bm{\Theta}(\mathbf{C},f_b)\mathbf{G}_b + \mathbf{h}_{bk}^H
    \right\|_2^2,
    \label{prob_rev_2a}\\[0mm]
    &\text{s.t.} \hspace{1.2mm}  \bm{\Theta}(\mathbf{C},f_b) = \mathrm{bdiag}(\bm{\Theta}_1(\mathbf{C}_1,f_b), \cdots, \bm{\Theta}_G(\mathbf{C}_G,f_b)), \label{prob_rev_2b}\\
    & \hspace{5.3mm} \bm{\Theta}_g(\mathbf{C}_g,f_b) \bm{\Theta}_g(\mathbf{C}_g,f_b)^H \preceq \mathbf{I}_{\bar{D}},\\ 
    & \hspace{5.3mm} \bm{\Theta}_g(\mathbf{C}_g,f_b) = \bm{\Theta}_g(\mathbf{C}_g,f_b)^T, \label{prob_rev_2c}
\end{align}
\end{subequations}
where $\mathbf{C}_g$ is the capacitance matrix associated with the $g$th RIS group, such that $\mathbf{C} = \mathrm{bdiag}(\mathbf{C}_1, \cdots, \mathbf{C}_G)$.

Due to the fact that $\bm{\Theta}$ is a function of both $\mathbf{C}$ and $f_b$, combined with the matrix constraints in \eqref{prob_1b} and \eqref{prob_1c}, and \eqref{prob_rev_2b}--\eqref{prob_rev_2c}, problems \eqref{prob_1} and \eqref{prob_rev_2} become intractable to solve.
Furthermore, as revealed in Section \ref{SecBDRIS}, in fully-connected RISs or within the groups of group-connected RISs, we cannot dedicate the practical reflecting elements for more than one operating frequency simultaneously given that elements are coupled with one another. Such characteristics further complicate their optimization process. Next, efficient relaxed strategies are proposed to tackle this challenge.

\subsection{Relaxed optimization for scenarios with obstructed direct links  for fully-connected RIS}\label{fRISopt}}
{We start by studying the scenario where the direct links between the BSs and users are under deep fading, meaning that the received power propagates dominantly through the reflected \ac{BS}-RIS-user links. In this case, we can neglect the effect of the direct channel vector $\mathbf{h}_{bk}$ in \eqref{prob_1a} and maximize the received power by matching only the channels between the BSs and the \ac{RIS} with the corresponding channels between the \ac{RIS} and users. More specifically, instead} 
%
%
%
of solving the original problem in \eqref{prob_1} directly, we apply a few relaxations to tackle the discussed challenges.
{To this end, we first formulate a relaxed optimization strategy} that considers the complex baseband version of the channels for all operating frequencies. Then, later, the achieved relaxed solution is integrated into a codebook-based strategy targeted at a single priority frequency, i.e., due to the frequency constraints of fully-connected RISs, to finally compute the practical reflecting coefficients. This practical configuration approach is explained in the following. First, we disregard the capacitance and frequency-dependent behavior of $\bm{\Theta}$, and, second, we relax constraint \eqref{prob_1b}. This simplified version of the problem can be expressed as follows:
\begin{subequations}\label{prob_2}
\begin{align}
    \underset{\bm{\Theta}}{\max} \hspace{3mm} &
     \sum_{b \in \mathcal{B}} \mu_{b} \sum_{k \in \mathcal{K}_b} \nu_{k} \left\| \mathbf{f}_{bk}^H \bm{\Theta}\mathbf{G}_b
    \right\|_2^2,
    \label{prob_2a}\\[0mm]
    \text{s.t.} \hspace{4mm}& \frac{1}{\sqrt{D}}\|\bm{\Theta}\|_F \leq 1, \label{prob_2b} \\
    \hspace{4mm} &  \bm{\Theta} = \bm{\Theta}^T \label{prob_2c}.
\end{align}
\end{subequations}

Problem \eqref{prob_2} is still challenging to solve due to the matrix constraints \eqref{prob_2b} and \eqref{prob_2c}. Fortunately, by applying a few transformations, we can achieve an equivalent convex version of such a problem, as follows. First, recall that the elements above the main diagonal of $\bm{\Theta}$ will be redundant due to its symmetric structure. This implies that the desired reflection coefficients can be optimized based on a non-redundant reduced-dimension version of $\bm{\Theta}$. Such a goal can be achieved with the help of Properties I and II, introduced next.

\vspace{1mm}

\noindent {\bf Property I:} Let $\mathbf{A}, \mathbf{B}$, and $\mathbf{C}$ represent arbitrary matrices of compatible dimensions. Then, the following Kronecker identity can be applied:
\begin{align}\label{kron_id}
    \mathrm{vec}(\mathbf{A}\mathbf{B}\mathbf{C}) = (\mathbf{C}^T \otimes \mathbf{A})\mathrm{vec}(\mathbf{B}).
\end{align}

\noindent {\bf Property II:} Let $\mathbf{A}$ be an arbitrary $D \times D$ symmetric matrix. Then, the following transformation holds:
\begin{align}
\mathbf{D}_D \mathrm{vech}(\mathbf{A}) = \mathrm{vec}(\mathbf{A}),    
\end{align}
where $\mathbf{D}_D$ is the duplication matrix of order $D$, which consists of a $D^2 \times D(D+1) / 2$ full column rank sparse matrix that can be uniquely defined by
\begin{align}\label{dup_mat}
    \mathbf{D}_D^T \triangleq \sum_{i=1}^{D} \sum_{j=1}^{i} \bm{u}_{ij} \mathrm{vec}(\mathbf{T}_{ij})^T,
\end{align}
where $\bm{u}_{ij}$ is a unit vector of length $D(D+1) / 2$ with its $[(j-1) D + i - j (j - 1)/2]$th entry equals to $1$ and zeros elsewhere, and $\mathbf{T}_{ij}$ is a $D\times D$ matrix with $1$ in its entries $(i j)$ and $(j i)$, and zeros elsewhere, for $1\leq j\leq i\leq D$.

\vspace{1mm}

\noindent {\it Proof:} Please, refer to \cite[Definitions 3.2a and 3.2b]{Magnus80}.\hfill \qedsymbol

\begin{figure}
\centering
{\begingroup
\csname @twocolumnfalse\endcsname
\resizebox{.47\textwidth}{!}{%
\begin{minipage}{.56\textwidth}
\setlength{\algomargin}{5mm}
\setlength{\interspacetitleboxruled}{1mm}
\begin{algorithm}[H]
\SetKwRepeat{Do}{do}{while}
\SetAlgoLined

\KwIn{$D;$ \vspace{1mm}}
\KwOut{$\mathbf{D}_{D};$ \vspace{2mm}}

Initialize: $\mathbf{D}^{(0)}_{D} = \mathbf{0}_{D^2, \frac{D (D + 1)}{2}}$\;\vspace{1mm}

\For{$i = 1, \cdots, D$}{\Indmm 
    \For{$j = 1, \cdots, i$}{\Indmm 
        \quad Generate the vector $\mathbf{u}_{ij}$ of length $\frac{D(D+1)}{2} \times 1$ in \eqref{dup_mat}\;\vspace{.5mm}
        
        \quad Generate the matrix $\mathbf{T}_{ij}$ with size $D\times D$ in \eqref{dup_mat}\;\vspace{.5mm}
        
        \quad Update the duplication matrix: \hfill $\left(\mathbf{D}^{\left(\frac{i (i - 1)}{2} + j \right)}_{D}\right)^T \hspace{-1mm} = \left(\mathbf{D}^{\left(\frac{i (i - 1)}{2} + j - 1\right)}_{D}\right)^T \hspace{-1mm} + \mathbf{u}_{ij} \mathrm{vec}(\mathbf{T}_{ij})^T$\;
}
}\vspace{-1mm}
Obtain the final duplication matrix:
$\mathbf{D}_{D} = \mathbf{D}^{\left(\frac{D (D + 1)}{2}\right)}_{D}$.

\caption{Generation of the duplication matrix of order $D$.}\label{alg_dmat}
\end{algorithm}
\end{minipage}
}%
\endgroup} 
\end{figure}

\vspace{1mm}

{The detailed procedure for computing the duplication matrix $\mathbf{D}_{D}$ is provided in Algorithm \ref{alg_dmat}.}
With the aid of Property II and exploiting the Kronecker identity in Property I, we can define
\begin{align}
    \mathbf{R}_{bk} &\triangleq \mathbf{G}_b^T \otimes \mathbf{f}_{bk}^H \in \mathbb{C}^{M \times D^2}\label{def_1}, \\
    \bm{\theta} &\triangleq \mathrm{vech}(\bm{\Theta}) \in \mathbb{C}^{\frac{D(D+1)}{2} \times 1}\label{def_2}.
\end{align}

Then, by plugging transformations \eqref{def_1} and \eqref{def_2} into \eqref{prob_2}, we can obtain a simplified equivalent version of the problem as follows
\begin{subequations}\label{prob_3}
\begin{align}
    \underset{\bm{\theta}}{\max} \hspace{3mm} &
     \sum_{b \in \mathcal{B}} \mu_{b} \sum_{k \in \mathcal{K}_b} \nu_{k} \left\| \mathbf{R}_{bk} \mathbf{D}_D\bm{\theta} 
    \right\|_2^2, \\[0mm]
    \text{s.t.} \hspace{4mm}& \frac{1}{\sqrt{D}}\|\mathbf{D}_D \bm{\theta}\|_2 \leq 1 \label{prob_3b}.
\end{align}
\end{subequations}

Problem \eqref{prob_3} can be further simplified by exploiting the following proposition.

\vspace{1mm}

\noindent {\bf Proposition I:} Let $\mathbf{D}_D$ denote a duplication matrix of order $D \geq 2$, computed through \eqref{dup_mat}, and $\bm{\theta} \in \mathbb{C}^{\frac{D(D+1)}{2}\times 1}$ denote a vector comprising the non-redundant lower triangular half of an arbitrary $D\times D$ symmetric matrix $\bm{\Theta}$, i.e., $\bm{\theta} = \mathrm{vech}(\bm{\Theta})$. Then, by meeting $\|\bm{\theta}\|_2 \leq 1$, the constraint $\frac{1}{\sqrt{D}}\|\mathbf{D}_D \bm{\theta}\|_2 \leq 1$ is satisfied with probability one.

\vspace{1mm}

\noindent {\it Proof:} Please, refer to Appendix \ref{ap1}.  \hfill \qedsymbol

\vspace{1mm}

\begin{figure}
\centering
\begingroup
\csname @twocolumnfalse\endcsname
\resizebox{.47\textwidth}{!}{%
{\begin{minipage}{.55\textwidth}
\setlength{\algomargin}{5mm}
\setlength{\interspacetitleboxruled}{1mm}
\begin{algorithm}[H]
\SetKwRepeat{Do}{do}{while}
\SetAlgoLined

\KwIn{$\mu_b, \nu_k, \mathbf{D}_{D}, \mathbf{G}_{b}, \mathbf{f}_{bk}, \forall k \in \mathcal{K}_b, \forall b \in \mathcal{B};$ \vspace{1mm}}
\KwOut{$\bm{\Theta}^*;$ \vspace{1mm}}

Construct:
$\mathbf{\hat{R}} = [\sqrt{\mu_1}\mathbf{\bar{R}}_{1}^T \cdots \sqrt{\mu_{B}}\mathbf{\bar{R}}_{B}^T]^T \mathbf{D}_D$, with: \vspace{1mm}

\nonl \hspace{2mm} $\mathbf{\bar{R}}_b = [\sqrt{\nu_{1}}\mathbf{R}_{b1}^T \cdots \sqrt{\nu_{K_b}}\mathbf{R}_{bK_b}^T]^T$; \vspace{.8mm}

\nonl \hspace{2mm} $\mathbf{R}_{bk} = \mathbf{G}_b^T \otimes \mathbf{f}_{bk}^H$;

\vspace{0.8mm}

Decompose the matrix $\mathbf{\hat{R}}$ through SVD: $\mathbf{\hat{R}} = \mathbf{U}\bm{\Lambda} \mathbf{V}^H$; \vspace{.8mm}

Compute the optimal non-redundant RIS coefficient vector by 

\nonl selecting the first eigenvector in $\mathbf{V}$: \vspace{.8mm}

\nonl \hspace{2mm} $\bm{\theta}^* = [\mathbf{V}]_{:1}  \in \mathbb{C}^{\frac{D(D+1)}{2} \times 1}$; \vspace{1mm}

Compute the relaxed reflection matrix:
$\bm{\Theta}^* = \mathrm{unvec}\{\mathbf{D}_{D} \bm{\theta}^*\}$.

\caption{Relaxed closed-form optimization for fully-connected RIS  when the direct BS-user link is not available.}\label{alg0}
\end{algorithm}
\end{minipage}}
}%
\endgroup \vspace{0mm}
\end{figure}

Supported by Proposition I, we can define
\begin{align}
\mathbf{\bar{R}}_b \triangleq [\sqrt{\nu_{1}}\mathbf{R}_{b1}^T \cdots \sqrt{\nu_{K_b}}\mathbf{R}_{bK_b}^T]^T \in \mathbb{C}^{K_b M \times D^2},\label{def_3}
\end{align}
and further relax problem \eqref{prob_3} so that the following can be achieved
\begin{subequations}\label{prob_4}
\begin{align}
        \underset{\bm{\theta}}{\max} \hspace{3mm} & \hspace{-.7mm}
        \left\| \renewcommand*{\arraystretch}{1} \setlength{\arraycolsep}{.5mm}
        \begin{bmatrix}
       \sqrt{\mu_1}\mathbf{\bar{R}}_{1} \\
       \vdots \\
       \sqrt{\mu_{B}}\mathbf{\bar{R}}_{B}
    \end{bmatrix} \mathbf{D}_D\bm{\theta}
         \right\|_2^2, \\
         \text{s.t.} \hspace{4mm} & \| \bm{\theta} \|_2 \leq 1,\label{prob_4b}
\end{align}
\end{subequations}
where the constraint in \eqref{prob_4b} still satisfies the law of conservation of energy, as demonstrated by Proposition I, which reflects the passive operation of the fully-connected RIS. The problem in \eqref{prob_4} is convex and can be solved by exploiting the SVD of the matrix $\mathbf{\hat{R}} \triangleq [\sqrt{\mu_1}\mathbf{\bar{R}}_{1}^T \cdots \sqrt{\mu_{B}}\mathbf{\bar{R}}_{B}^T]^T \mathbf{D}_D \in \mathbb{C}^{K_b M B \times \frac{D(D+1)}{2}}$. Specifically, by recalling the SVD, we can write $\mathbf{\hat{R}} = \mathbf{U}\bm{\Lambda} \mathbf{V}^H$. Then, the solution of \eqref{prob_4} can be computed as $\bm{\theta}^* = [\mathbf{V}]_{:1}$, which consists of the eigenvector associated with the largest eigenvalue of the transformed stacked matrix $\mathbf{\hat{R}}$. Thus, the relaxed optimal \ac{RIS} scattering matrix can be computed as $\bm{\Theta}^* = \mathrm{unvec}(\mathbf{D}_D\bm{\theta}^*)$, with $\mathbf{D}_D$ given in \eqref{dup_mat}. The steps for the computation of the closed-form solution of \eqref{prob_4} are summarized in Algorithm \ref{alg0}.

{\subsection{Relaxed optimization for scenarios with unobstructed direct links for fully-connected RIS}\label{relax_fris_subsec}
}
{We have derived in the previous subsection a closed-form solution for the relaxed problem in \eqref{prob_4}, which is valid for scenarios where the BS transmissions reach users only through RIS reflected channel links. In this subsection, we generalize our optimization strategy to the case in which signal reception in the direct BS-user link is non-negligible. Specifically, in a scenario with unobstructed direct links, our relaxed objective for the fully-connected RIS can be formulated as follows}
{\begin{subequations}\label{prob_rev_1}
\begin{align}
    \underset{\bm{\Theta}}{\max} \hspace{3mm} &
     \sum_{b \in \mathcal{B}} \mu_{b} \sum_{k \in \mathcal{K}_b} \nu_{k} \left\| \mathbf{f}_{bk}^H \bm{\Theta}\mathbf{G}_b + \mathbf{h}_{bk}^H
    \right\|_2^2,
    \label{prob_rev_1a}\\[0mm]
    \text{s.t.} \hspace{4mm}& \frac{1}{\sqrt{D}}\|\bm{\Theta}\|_F \leq 1, \label{prob_rev_1b} \\
    \hspace{4mm} &  \bm{\Theta} = \bm{\Theta}^T \label{prob_rev_1c},
\end{align}
\end{subequations}
where we have reintroduced the channel vector $\mathbf{h}_{bk}$ in the objective function in \eqref{prob_rev_1a}, and the same relaxations employed in the previous subsection have been applied. To solve problem \eqref{prob_rev_1}, we first rely on Property I and further define}
{\begin{align}
    \mathbf{\underline{h}}_{bk} &\triangleq \mathrm{vec}(\mathbf{h}_{bk}^H) \in \mathbb{C}^{M \times 1}\label{def_4}.
\end{align}
Then, by recalling Property II and Proposition I, and applying the transformations in \eqref{def_1}, \eqref{def_2}, and \eqref{def_4}, we can achieve
\begin{subequations}\label{prob_rev_3}
\begin{align}
    \underset{\bm{\theta}}{\max} \hspace{3mm} &
     \sum_{b \in \mathcal{B}} \mu_{b} \sum_{k \in \mathcal{K}_b} \nu_{k} \left\| \mathbf{R}_{bk} \mathbf{D}_D\bm{\theta} + \mathbf{\underline{h}}_{bk}
    \right\|_2^2, \\[0mm]
    \text{s.t.} \hspace{4mm}& \|\bm{\theta}\|_2 \leq 1 \label{prob_rev_3b}.
\end{align}
\end{subequations}

As the last step, we introduce the notation $\mathbf{\bar{h}}_{b} \triangleq [\sqrt{\nu_{1}}\mathbf{\underline{h}}_{b1}^T, \cdots, \sqrt{\nu_{1}}\mathbf{\underline{h}}_{bK_b}^T]^T$ and recall the definition in \eqref{def_3}, which allow us to apply a stacking strategy to simplify \eqref{prob_rev_3} into its final form, as follows
\begin{subequations}\label{prob_rev_4}
\begin{align}
        \underset{\bm{\theta}}{\max} \hspace{3mm} & \hspace{-.7mm}
        \left\| \renewcommand*{\arraystretch}{1} \setlength{\arraycolsep}{.5mm}
        \begin{bmatrix}
       \sqrt{\mu_1}\mathbf{\bar{R}}_{1} \\
       \vdots \\
       \sqrt{\mu_{B}}\mathbf{\bar{R}}_{B}
    \end{bmatrix} \mathbf{D}_D\bm{\theta}
         + \begin{bmatrix}
             \sqrt{\mu_1}\mathbf{\bar{h}}_{1} \\
             \vdots \\
             \sqrt{\mu_1}\mathbf{\bar{h}}_{B}
         \end{bmatrix} \right\|_2^2, \label{prob_rev_4a}\\
         \text{s.t.} \hspace{4mm} & \| \bm{\theta} \|_2 \leq 1.\label{prob_rev_4b}
\end{align}
\end{subequations}
In contrast to the problem \eqref{prob_4}, a closed-form optimal solution for \eqref{prob_rev_4} cannot be achieved. Nevertheless, given that the objective function \eqref{prob_rev_4a} is convex and continuously differentiable, and that the constraint set is the closed unit-radius ball of the $L_2$ norm, we can exploit the conditional gradient method \cite{Liu2020} to efficiently solve problem \eqref{prob_rev_4}. The proposed optimization strategy is presented in Algorithm \ref{alg_rev_1}. As demonstrated in \cite{Jaggi21}, when optimizing over an $L_2$-ball domain, the iterates $\bm{\theta}^{(i)}$ of Algorithm \ref{alg_rev_1} satisfy $f(\bm{\theta}^{(i)}) - f(\bm{\theta}^*) = \mathcal{O}(\frac{1}{i})$, with $f(\cdot)$ denoting the given objective function, implying that the iterates converge to the global optimal solution $\bm{\theta}^*$ at a rate of $\mathcal{O}(\frac{1}{i})$, with $i=1,\cdots, I$, as the number of iterations $I \rightarrow \infty$.}

\begin{figure}
\centering
\begingroup
\csname @twocolumnfalse\endcsname
\resizebox{.47\textwidth}{!}{%
{\begin{minipage}{.56\textwidth}
\setlength{\algomargin}{5mm}
\setlength{\interspacetitleboxruled}{1mm}
\begin{algorithm}[H]
	\SetKwRepeat{Do}{do}{while}
	\SetAlgoLined
	\KwIn{\small $I, \mu_b, \nu_k, \mathbf{D}_{D}, \mathbf{G}_{b}, \mathbf{f}_{bk}, \mathbf{h}_{bk}, \forall k \in \mathcal{K}_b, \forall b \in \mathcal{B}$\;\vspace{1mm}}
    \KwOut{$\bm{\Theta}^*$;\vspace{1mm}}

	Construct:  $\mathbf{\hat{h}} = [\sqrt{\mu_1}\mathbf{\bar{h}}_{1}^T \cdots \sqrt{\mu_{B}}\mathbf{\bar{h}}_{B}^T]^T$, and
 
\nonl $\mathbf{\hat{R}} = [\sqrt{\mu_1}\mathbf{\bar{R}}_{1}^T \cdots \sqrt{\mu_{B}}\mathbf{\bar{R}}_{B}^T]^T \mathbf{D}_D$, with: \vspace{1mm}

\nonl \hspace{3mm} $\mathbf{\bar{h}}_{b} = [\sqrt{\nu_{1}}\mathbf{\underline{h}}_{b1}^T, \cdots, \sqrt{\nu_{K_b}}\mathbf{\underline{h}}_{bK_b}^T]^T$; \vspace{.8mm}

\nonl \hspace{3mm} $\mathbf{\bar{R}}_b = [\sqrt{\nu_{1}}\mathbf{R}_{b1}^T \cdots \sqrt{\nu_{K_b}}\mathbf{R}_{bK_b}^T]^T$; \vspace{.8mm}

\nonl \hspace{3mm} $\mathbf{\underline{h}}_{bk} = \mathrm{vec}(\mathbf{h}_{bk}^H)$; \vspace{.8mm}

\nonl \hspace{3mm} $\mathbf{R}_{bk} = \mathbf{G}_b^T \otimes \mathbf{f}_{bk}^H$; \vspace{.8mm}

	Initialize $\bm{\theta}^{(1)}$ with an arbitrary vector in the feasible set:
    $\bm{\theta}^{(1)} \in \{\mathbf{x} : \|\mathbf{x}\|^2 \leq 1, \mathbf{x} \in \mathbb{C}^{\frac{D(D+1)}{2} \times 1} \}$\;\vspace{1mm}
        
		\For{$i = 1, 2, \cdots, I - 1$}{\Indmm 
		Compute the gradient of the objective function in \eqref{prob_rev_4a}: \hspace{50mm}
        $\bm{\vartheta}^{(i)} = 2 \mathbf{\hat{R}}^H (\mathbf{\hat{R}} \bm{\theta}^{(i)} + \mathbf{\hat{h}})$\;\vspace{1mm}
				
		Solve the direction-finding sub-problem:
  
        \nonl \hspace{1mm} $\mathbf{s}^{(i)} = \arg\min -(\mathbf{s}^{(i)})^T \bm{\vartheta}^{(i)} \hspace{1mm} \text{ s.t. } \hspace{1mm} \|\mathbf{s}^{(i)}\|_2\leq 1$:\vspace{1mm}

        {\Indpp 
		\For{$l = 1, 2, \cdots, D(D+1)/2$}{\vspace{1mm}
		$[\mathbf{s}^{(i)}]_{l} = \mathrm{sign}([\bm{\vartheta}^{(i)}]_l) \frac{|[\bm{\vartheta}^{(i)}]_l|}{\|\bm{\vartheta}^{(i)}\|_2};$
		}
  \vspace{1mm}}

        Update the step size: $\xi^{(i)} = 2/(i + 2)$\;

        Update the reflection coefficient vector: $\bm{\theta}^{(i+1)} = (1-\xi^{(i)})\bm{\theta}^{(i)} + \xi^{(i)} \mathbf{s}^{(i)};$
        
        }
		
        Compute the relaxed reflection matrix: $\bm{\Theta}^* = \mathrm{unvec}\{\mathbf{D}_{D} \bm{\theta}^{(I)}\}.$
	\caption{Relaxed conditional gradient-based strategy for fully-connected RIS when the direct BS-user link is available.}\label{alg_rev_1}
\end{algorithm}
\end{minipage}
}}%
\endgroup
\end{figure}

\vspace{2mm}
{\subsection{Codebook-based practical frequency-aware configuration for fully-connected RIS}\label{subsec_cb_fc}
}

Recall that we are interested in a matrix with practical capacitance values to configure the presented \ac{RIS} architecture. The work in \cite{Cai21}, for instance, studied a multi-band scenario assisted by a conventional single-connected \ac{RIS} and proposed a strategy of dedicating each reflecting coefficient to one distinct frequency. However, optimizing each coefficient individually is not applicable to the fully-connected \ac{RIS} case due to the interconnected elements. As previously explained, this implies that a practical fully-connected frequency-dependent \ac{RIS} needs to be configured based on a priority frequency $f^\star$. To this end, we propose next a codebook-based strategy for mapping, i.e., approximating, the optimal matrix $\bm{\Theta}^*${, i.e., computed either through Algorithm \ref{alg0} or Algorithm \ref{alg_rev_1},} to the desired practical solution targeted at frequency $f^\star$.{
To accomplish this, we start by computing the corresponding relaxed impedance matrix with the aid of Lemma I, which is presented next. 

\vspace{1mm}

\noindent {\bf Lemma I:} Given a $D \times D$ scattering matrix $\bm{\Theta}$ for a fully-connected RIS, its associated impedance matrix $\mathbf{Z}$ can be calculated as follows:
\begin{align}\label{impedmat1}
      \mathbf{Z} = Z_0 (\mathbf{I}_{D} + \bm{\Theta}) (\mathbf{I}_{D} -  \bm{\Theta})^{-1}.
\end{align}

\vspace{1mm}

\noindent {\it Proof:} Please, refer to Appendix \ref{ap2}. \hfill \qedsymbol

}

\vspace{1mm}

Let $\mathbf{Z}^*$ denote the relaxed symmetric transfer impedance matrix computed based on $\bm{\Theta}^*$ through Lemma I. Then, by exploiting the definition in \eqref{admit_mat}, the non-redundant relaxed self and inter-element impedances associated with $\mathbf{Z}^*$, can be retrieved by
\begin{align}
    \tilde{Z}^*_{pq} &= - \frac{1}{[(\mathbf{Z}^*)^{-1}]_{pq}}, \quad \forall p < q, \label{relax_imp_1}\\[1mm]
    Z^*_{p} &= \frac{1}{\sum_{i=1}^D [(\mathbf{Z}^*)^{-1}]_{pi}}, \quad \forall p = q.\label{relax_imp_2}
\end{align}

Note that the values of $Z^*_{p}$ and $\tilde{Z}^*_{pq}$ can assume infinitely arbitrary values due to the fact that the constraint \eqref{prob_1b} has been relaxed. To model the practical behavior of the physical circuit components, we construct two codebooks, namely $\mathcal{Z} = \{Z_{i}(C_{i}, f^\star)\}$ and $\mathcal{\tilde{Z}} = \{\tilde{Z}_{i}(\tilde{C}_{i}, f^\star)\}$, which comprise the frequency-dependent self and inter-element impedances, $Z_{i}(C_{i}, f^\star)$ and $\tilde{Z}_{i}(\tilde{C}_{i}, f^\star)$, generated through \eqref{selfimp_eq} and \eqref{intelimp_eq}, respectively, in which $i = 1, \cdots, 2^{B_{C}}$, with $B_{C}$ denoting the number of capacitance quantization bits used in the two codebooks. After obtaining the codebooks, the optimized practical impedance connecting the elements $p$ and $q$, $\forall p < q$, with $p, q \in \{1, \cdots, D\}$, corresponding to the target frequency $f^\star$, can be determined by
\begin{align}\label{prob_5}
    \hat{Z}^*_{pq}(\hat{C}^*_{pq}, f^\star) & = \underset{\forall \tilde{\zeta}_i \in \mathcal{\tilde{Z}} | i = 1, \cdots,2^{B_{C}}}{\arg \min } \hspace{2mm}
     | \tilde{Z}^*_{pq} - \tilde{\zeta}_i
    |^2,
\end{align}
while the desired $p$th practical self-impedance, $\forall p \in \{1, \cdots, D\}$, can be obtained as
\begin{align}\label{prob_5_2_rev}
    \hat{Z}^*_{p}(\hat{C}^*_{p}, f^\star) & = \underset{\forall \zeta_i \in \mathcal{Z} | i = 1, \cdots,2^{B_{C}}}{\arg \min } \hspace{2mm}
     | Z^*_{p} - \zeta_i
    |^2.
\end{align}

Once the practical impedances are obtained with \eqref{prob_5} and \eqref{prob_5_2_rev}, we can recall \eqref{admit_mat} and promptly compute the corresponding practical transfer impedance matrix, which is then used to determine $\bm{\Theta}^*(\mathbf{\hat{C}}^*, f^\star)$ with the help of \eqref{scatmat1}. 
{Note that we have a one-to-one mapping between the practical impedances obtained through \eqref{prob_5} and \eqref{prob_5_2_rev} and the matrix with the desired reconfigurable capacitance values $\mathbf{\hat{C}}^*$. 
After determining $\mathbf{\hat{C}}^*$}, we can compute the scattering matrices $\bm{\Theta}(\mathbf{\hat{C}}^*, f_b)$, through \eqref{scatmat1}--\eqref{intelimp_eq}, for the remaining operating frequencies $f_b \in \mathcal{F}$.
The proposed practical configuration strategy for the fully-connected RIS is summarized in Algorithm \ref{alg1}.

\begin{figure}
\centering
\begingroup
\csname @twocolumnfalse\endcsname
\resizebox{.47\textwidth}{!}{%
\begin{minipage}{.56\textwidth}
\setlength{\algomargin}{5mm}
\setlength{\interspacetitleboxruled}{1mm}
\begin{algorithm}[H]
\SetKwRepeat{Do}{do}{while}
\SetAlgoLined

\KwIn{$\mu_b, \nu_k, D, f^\star, \mathcal{Z}, \mathcal{\tilde{Z}}, \mathbf{G}_{b}, \mathbf{f}_{bk}, \mathbf{h}_{bk}, \forall k \in \mathcal{K}_b, \forall b \in \mathcal{B};$ \vspace{1mm}}
\KwOut{$\mathbf{C}^*;$ \vspace{1mm}}

Providing $D$, generate the duplication matrix $\mathbf{D}_{D}$ with Algorithm \ref{alg_dmat};

\vspace{1mm}

{
Use $\mathbf{D}_{D}$ and the listed input parameters to obtain the

\nonl relaxed RIS reflection matrix $\bm{\Theta}^*$:

\vspace{1mm}

\Indp
\uIf{direct BS-user channels are blocked}{
    With Algorithm \ref{alg0};
}\vspace{1mm}

\uIf{direct BS-user channels are available}{
    With Algorithm \ref{alg_rev_1};
}
}

\vspace{1mm}

Obtain the corresponding relaxed impedance matrix: $\mathbf{Z}^* = Z_0 (\mathbf{I}_{D} + \bm{\Theta}^*) (\mathbf{I}_{D} -  \bm{\Theta}^*)^{-1}$;
\vspace{0.5mm}

{Retrieve the non-redundant self and inter-element impedances, $Z^*_{p}$ and $\tilde{Z}^*_{pq}$, associated with $\mathbf{Z}^*$, using \eqref{relax_imp_1} and \eqref{relax_imp_2};}

\vspace{1mm}

{Determine the practical impedances $\hat{Z}^*_{p}(\hat{C}^*_{p}, f^\star)$ and $\hat{Z}^*_{pq}(\hat{C}^*_{pq}, f^\star)$, for the target frequency $f^\star \in \{f_1, \cdots, f_B\}$, by exploiting the codebooks $\mathcal{Z}$ and $\mathcal{\tilde{Z}}$ in \eqref{prob_5} and \eqref{prob_5_2_rev};}

\vspace{1mm}

{With $\hat{Z}^*_{p}(\hat{C}^*_{p}, f^\star)$ and $\hat{Z}^*_{pq}(\hat{C}^*_{pq}, f^\star)$,} finally obtain the matrix with the practical capacitance values $\mathbf{\hat{C}}^*$.

\caption{Frequency-aware configuration strategy for fully-connected RIS.}\label{alg1}
\end{algorithm}
\end{minipage}
}%
\endgroup 
\end{figure}

\vspace{3mm}
{\subsection{Relaxed optimization for scenarios with  obstructed direct links for group-connected RIS}\label{gRISopt}}

As explained in Section \ref{SecBDRIS}, the groups of reflecting elements in a group-connected \ac{RIS} are independent. This characteristic allows us to assign each group to a distinct operating frequency, a capability that is not possible with a fully-connected \ac{RIS}. In this subsection, we consider the scenario with obstructed direct links and we optimize the reflecting coefficients based first on a relaxed version of problem \eqref{prob_rev_2}. In the second step of the optimization process, we will map the relaxed solution to practical frequency-dependent values based on a codebook-based approach{, similarly to the strategy employed in the previous subsection}.
The base-band relaxed optimization problem for the group-connected \ac{RIS} case{, neglecting the direct channels,} can be formulated as follows:
\begin{subequations}\label{prob_6}
\begin{align}
    \underset{\bm{\Theta}}{\max} \hspace{3mm}&
     \sum_{b \in \mathcal{B}} \mu_{b} \sum_{k \in \mathcal{K}_b} \nu_{k} \left\| \mathbf{f}_{bk}^H \bm{\Theta}\mathbf{G}_b
    \right\|_2^2,
    \label{prob_6a}\\[0mm]
    \text{s.t.} \hspace{4mm} & \bm{\Theta} = \mathrm{bdiag}(\bm{\Theta}_1, \cdots, \bm{\Theta}_G), \label{prob_6b}\\
    & \frac{1}{\sqrt{\Bar{D}}}\|\bm{\Theta}_g\|_F \leq 1,\\ 
    &\bm{\Theta}_g = \bm{\Theta}_g^T, \label{prob_6c}
\end{align}
\end{subequations}
where $\mu_{b}$ and $\nu_{k}$ are the optimization weights defined as in \eqref{prob_1}. Before we can solve \eqref{prob_6}, we need to apply a few transformations to overcome the challenging matrix constraints in \eqref{prob_6b}--\eqref{prob_6c}. First, we define
\begin{align}
    \mathbf{f}_{bkg} &= [\mathbf{f}_{bk}]_{(1 + (g-1) \Bar{D}) : g \Bar{D}},\label{def_gc_1}\\
    \mathbf{G}_{bg} &= [\mathbf{G}_b]_{(1 + (g-1) \Bar{D}) : g \Bar{D},:},\label{def_gc_2}
\end{align}
for $g = 1, \cdots, G$. Next, by plugging the above definitions into \eqref{prob_6}, we achieve
\begin{subequations}
\begin{align}
    \underset{\bm{\Theta}_1, \cdots, \bm{\Theta}_G}{\max} \hspace{2mm}& 
        \sum_{b \in \mathcal{B}} \mu_{b} \sum_{k \in \mathcal{K}_b} \nu_{k} \left\|\sum_{g = 1}^{G} \mathbf{f}_{bkg}^H \bm{\Theta}_g \mathbf{G}_{bg} \right\|_2^2,
    \\[0mm]
    \text{s.t.}  \hspace{6mm} &
     \frac{1}{\sqrt{\bar{D}}}\|\bm{\Theta}_g\|_F \leq 1 \\
     &\bm{\Theta}_g = \bm{\Theta}_g^T.
\end{align}
\end{subequations}
With the help of Properties I and II, we apply the following transformations
\begin{align}
    \mathbf{R}_{bkg} &\triangleq \mathbf{G}_{bg}^T \otimes \mathbf{f}_{bkg}^H \in \mathbb{C}^{M \times \Bar{D}^2}, \label{trans20}\\
    \bm{\theta}_g &\triangleq \mathrm{vech}(\bm{\Theta}_g) \in \mathbb{C}^{\frac{\Bar{D}(\Bar{D}+1)}{2} \times 1},
\end{align}
and define
\begin{align}
    \mathbf{\bar{R}}_{bk} &\triangleq [\mathbf{R}_{bk1}\mathbf{D}_{\Bar{D}} \hspace{2mm} \cdots \hspace{2mm} \mathbf{R}_{bkG}\mathbf{D}_{\Bar{D}} ],\label{def_gris_0}\\
    \bm{\bar{\theta}} &\triangleq [\bm{\theta}_1^T, \cdots, \bm{\theta}_G^T ]^T,\label{def_gris_01}
\end{align}
where $\mathbf{D}_{\Bar{D}}$ is the duplication matrix introduced in \eqref{dup_mat}. Then, by invoking \eqref{trans20}--\eqref{def_gris_01}, and recalling Proposition I, we can achieve the following problem
\begin{subequations}\label{prob_rex_gRIS}
\begin{align}
    \underset{\bm{\theta}}{\max} \hspace{3mm}& 
        \sum_{b \in \mathcal{B}} \mu_{b} \sum_{k \in \mathcal{K}_b} \nu_{k} \left\|\mathbf{\bar{R}}_{bk} \bm{\bar{\theta}} \right\|_2^2,\\[0mm]
    \text{s.t.} \hspace{4mm}& \frac{1}{\sqrt{G}}\|{\bm{\bar{\theta}}}\|_2 \leq 1.
\end{align}
\end{subequations}

Next, by exploiting the property that each RIS group is independent, we select $S \leq G$ priority BSs, which are represented by $\mathcal{S} = \{1, \cdots, S\} \subseteq \mathcal{B}$, such that $S \leq B$, to carry out the optimization. This allows us to assign one or more independent RIS groups to a distinct BS. To be specific, we select a subset $\mathcal{G}_s  = \{1, \cdots, G_s\}$ comprising $G_s$ RIS groups, in which $1 \leq G_s \leq G$, to assist the $s$th BS, where we must ensure that the group subsets are disjoint, i.e., $\mathcal{G}_s \cap \mathcal{G}_{s'} = \text{\O}, \forall s \neq s'$, implying that $\sum_{s=1}^{S} G_s = G$. To this end, we define $\mathbf{\tilde{R}}_s \triangleq \sqrt{\mu_{s}}[\sqrt{\nu_1}\mathbf{\bar{R}}_{s1}^T \cdots \sqrt{\nu_{K_s}}\mathbf{\bar{R}}_{sK_s}^T]^T$, with $s \in \mathcal{S} \subseteq \mathcal{B}$, and decouple the problem in \eqref{prob_rex_gRIS} into $S$ independent sub-problems, as follows
\begin{subequations}\label{prob_rex_gRIS2}
\begin{align}
    \underset{{\bm{\tilde{\theta}}_s}}{\max} \hspace{3mm} & 
         \left\|\mathbf{\tilde{R}}_{s} \bm{\tilde{\theta}}_s \right\|_2^2,\\[0mm]
    \text{s.t.} \hspace{4mm}& \frac{1}{\sqrt{G}}\|{\bm{\tilde{\theta}}_s}\|_2 \leq 1,
\end{align}
\end{subequations}
where ${\bm{\tilde{\theta}}_s}$ represents the non-redundant RIS coefficient vector for the $s$th priority BS, $\forall s \in \mathcal{S} \subseteq \mathcal{B}$. The solution for the $s$th problem in \eqref{prob_rex_gRIS2} can be obtained with the aid of the SVD, similarly as for the problem \eqref{prob_4}. More specifically, we can decompose $\mathbf{\tilde{R}}_s = \mathbf{\tilde{U}}_s\bm{\tilde{\Lambda}}_s \mathbf{\tilde{V}}_s^H$ and achieve the desired solution by computing ${\bm{\tilde{\theta}}^*_s} = \sqrt{G} [\mathbf{\tilde{V}}_s]_{:1}$.
Finally, the relaxed coefficients for each independent \ac{RIS} group for the $s$th BS can be obtained as $\bm{\theta}^*_{sg} = [{\bm{\tilde{\theta}}_s^*}]_{(1 + (g-1) \Bar{D}): g \Bar{D}}$, and $\bm{\Theta}^*_{sg} = \mathrm{unvec}(\mathbf{D}_{\Bar{D}}\bm{\theta}^*_{sg})$, for $g \in \mathcal{G}_s$. The summary of the relaxed closed-form optimization for the group-connected RIS is presented in Algorithm \ref{alg_rev_0}.

\begin{figure}
\centering
\begingroup
\csname @twocolumnfalse\endcsname
\resizebox{.47\textwidth}{!}{%
\begin{minipage}{.55\textwidth}
\setlength{\algomargin}{5mm}
\setlength{\interspacetitleboxruled}{1mm}
{\begin{algorithm}[H]
\SetKwRepeat{Do}{do}{while}
\SetAlgoLined

\KwIn{
$\mu_s, \nu_k, \Bar{D}, G, \mathcal{G}_s, \mathbf{D}_{\Bar{D}}, \mathbf{G}_{s}, \mathbf{f}_{sk},$ $\forall k \in \mathcal{K}_s,$ $\forall s \in \mathcal{S}$\;\vspace{1mm}}
\KwOut{$\{\bm{\Theta}^*_{s1}, \cdots, \bm{\Theta}^*_{sG_s}\}$, $\forall s \in \mathcal{S}$\;}
 \vspace{1mm}

\For{\normalfont{\bf each} $s \in \mathcal{S}$}{\Indmm 

Construct: $\mathbf{\tilde{R}}_s = \sqrt{\mu_{s}}[\sqrt{\nu_1}\mathbf{\bar{R}}_{s1}^T \cdots \sqrt{\nu_{K_s}}\mathbf{\bar{R}}_{sK_s}^T]^T$,
 with: \vspace{1mm}

\nonl \hspace{2mm}$\mathbf{\bar{R}}_{sk} = [\mathbf{R}_{sk1}\mathbf{D}_{\Bar{D}} \hspace{1mm} \cdots \hspace{1mm} \mathbf{R}_{skG}\mathbf{D}_{\Bar{D}} ]$\;

\nonl \hspace{2mm}$\mathbf{R}_{skg} = \mathbf{G}_{sg}^T \otimes \mathbf{f}_{skg}^H$\;

\nonl \hspace{2mm}$\mathbf{f}_{skg} = [\mathbf{f}_{sk}]_{(1 + (g-1) \Bar{D}) : g \Bar{D}}$ \;

\nonl \hspace{2mm}$\mathbf{G}_{sg} = [\mathbf{G}_s]_{(1 + (g-1) \Bar{D}) : g \Bar{D},:}$\;

\vspace{1mm}

Decompose the matrix $\mathbf{\tilde{R}}_s$ through SVD: $\mathbf{\tilde{R}}_s = \mathbf{\tilde{U}}_s\bm{\tilde{\Lambda}}_s \mathbf{\tilde{V}}_s^H$\;

\vspace{1mm}

Compute the optimal coefficient vector for the $s$th BS by 

\nonl selecting the first eigenvector in $\mathbf{\tilde{V}}_s$: \vspace{1mm}

\nonl \hspace{2mm} $\bm{\tilde{\theta}}^*_s = \sqrt{G} [\mathbf{\tilde{V}}_s]_{:1}  \in \mathbb{C}^{G\frac{\bar{D}(\bar{D}+1)}{2} \times 1}$; \vspace{1mm}

\For{\normalfont{\bf each} $g \in \mathcal{G}_s$}{\Indmm
\hspace{1mm} Compute the sub-vector for the $g$th group: $\bm{\theta}^*_{sg} = [\bm{\tilde{\theta}}^*_s]_{(1 + (g-1) \Bar{D}): g \Bar{D}}$\;

\vspace{1mm}

\hspace{1mm} Compute the reflection block matrix for the $g$th group: $\bm{\Theta}^*_{sg} = \mathrm{unvec}(\mathbf{D}_{\Bar{D}}\bm{\theta}^*_{sg})$\;

}
}

\caption{Relaxed closed-form optimization for group-connected RIS when the direct BS-user link is not available.}\label{alg_rev_0}
\end{algorithm}}
\end{minipage}
}%
\endgroup \vspace{-2mm}
\end{figure}

\vspace{1mm}
{\subsection{Relaxed optimization for scenarios with unobstructed direct links for group-connected RIS}
\vspace{-1mm}
The relaxed optimization strategy for the group-connected RIS is now extended to the case where direct BS-user links are available. To optimize the group-connected RIS to such scenarios, we need to incorporate the contribution of the direct channel vector $\mathbf{h}_{bk}$ into \eqref{prob_6}, resulting in the following generalized problem:
\begin{subequations}\label{prob_rev_6}
\begin{align}
    \underset{\bm{\Theta}}{\max} \hspace{3mm}&
     \sum_{b \in \mathcal{B}} \mu_{b} \sum_{k \in \mathcal{K}_b} \nu_{k} \left\| \mathbf{f}_{bk}^H \bm{\Theta}\mathbf{G}_b + \mathbf{h}_{bk}^H
    \right\|_2^2,
    \label{prob_rev_6a}\\[0mm]
    \text{s.t.} \hspace{4mm} & \bm{\Theta} = \mathrm{bdiag}(\bm{\Theta}_1, \cdots, \bm{\Theta}_G), \label{prob_rev_6b}\\
    & \frac{1}{\sqrt{\Bar{D}}}\|\bm{\Theta}_g\|_F \leq 1.\\ 
    &\bm{\Theta}_g = \bm{\Theta}_g^T, \label{prob_rev_6c}
\end{align}
\end{subequations}}\vspace{-3mm}

{Before solving \eqref{prob_rev_6}, we first apply a few simplification procedures. We start by recalling the definitions of the sub-channels corresponding to each RIS group in \eqref{def_gc_1} and \eqref{def_gc_2}, which allows us to write
\begin{subequations}
\begin{align}
    \underset{\bm{\Theta}_1, \cdots, \bm{\Theta}_G}{\max} \hspace{2mm}& 
        \sum_{b \in \mathcal{B}} \mu_{b} \sum_{k \in \mathcal{K}_b} \nu_{k} \left\|\sum_{g = 1}^{G} \mathbf{f}_{bkg}^H \bm{\Theta}_g \mathbf{G}_{bg}  + \mathbf{h}_{bk}^H \right\|_2^2,
    \\[0mm]
    \text{s.t.}  \hspace{6mm} &
     \frac{1}{\sqrt{\bar{D}}}\|\bm{\Theta}_g\|_F \leq 1 \\
     &\bm{\Theta}_g = \bm{\Theta}_g^T.
\end{align}
\end{subequations}
Next, by invoking Properties I and II, and Proposition I, the above problem can be simplified as follows
\begin{subequations}\label{prob_rex_gRIS_rev}
\begin{align}
    \underset{\bm{\theta}}{\max} \hspace{3mm}& 
        \sum_{b \in \mathcal{B}} \mu_{b} \sum_{k \in \mathcal{K}_b} \nu_{k} \left\|\mathbf{\bar{R}}_{bk} \bm{\bar{\theta}} + \mathbf{\underline{h}}_{bk} \right\|_2^2,\\[0mm]
    \text{s.t.} \hspace{4mm}& \frac{1}{\sqrt{G}}\|\bm{\bar{\theta}}\|_2 \leq 1,
\end{align}
\end{subequations}
where the vector $\mathbf{\underline{h}}_{bk}$ is computed as in \eqref{def_4}, and $\mathbf{\bar{R}}_{bk}$ and $\bm{\bar{\theta}}$ are defined as in \eqref{def_gris_0} and \eqref{def_gris_01}, respectively.}

{As in Subsection \ref{gRISopt}, we rely on the fact that the RIS groups are independent and also select $S \leq G$ priority BSs, with $S \leq B$. Then, we dedicate one or more groups of the RIS, i.e., a subset $\mathcal{G}_s  = \{1, \cdots, G_s\}$ with $G_s$ independent groups, to assist a different BS $s \in \mathcal{S} \subseteq \mathcal{B}$, in which  $\sum_{s=1}^{S} G_s = G$ and $\mathcal{G}_s \cap \mathcal{G}_{s'} = \text{\O}, \forall s \neq s'$ must hold. For this, we transform \eqref{prob_rex_gRIS_rev} into the following decoupled sub-problems
\begin{subequations}\label{prob_rex_gRIS2_rev}
\begin{align}
    \underset{\bm{\tilde{\theta}}_s}{\max} \hspace{3mm} & 
         \left\|\mathbf{\tilde{R}}_{s} \bm{\tilde{\theta}}_s + \mathbf{\tilde{h}}_{s} \right\|_2^2,\label{prob_rex_gRIS2_a_rev}\\[0mm]
    \text{s.t.} \hspace{4mm}& \frac{1}{\sqrt{G}}\|\bm{\tilde{\theta}}_s\|_2 \leq 1,
\end{align}
\end{subequations}
where $\mathbf{\tilde{h}}_{s} \triangleq \sqrt{\mu_{s}}[\sqrt{\nu_{1}}\mathbf{\underline{h}}_{s1}^T, \cdots, \sqrt{\nu_{K_s}}\mathbf{\underline{h}}_{sK_s}^T]^T$ have been introduced, and $\mathbf{\tilde{R}}_s$ and $\bm{\tilde{\theta}}_s$ are defined as in \eqref{prob_rex_gRIS2}.}
{It is straightforward to show that each $s$th sub-problem in \eqref{prob_rex_gRIS2_rev} is convex. However, their global optimal solution cannot be computed in closed form. Therefore, as for problem \eqref{prob_rev_4}, we propose a conditional gradient-based approach to tackle problem \eqref{prob_rex_gRIS2_rev}. The detailed solution is presented in Algorithm \ref{alg_rev_2}. Since the objective function in \eqref{prob_rex_gRIS2_a_rev} is continuously differentiable, and the feasible set is defined by the $L_2$-norm ball of radius $\sqrt{G}$, the proposed method has a global convergence guaranteed at a rate of $\mathcal{O}(\frac{1}{i})$, as discussed in Subsection \ref{relax_fris_subsec}.}

\begin{figure}
\centering
\begingroup
\csname @twocolumnfalse\endcsname
\resizebox{.47\textwidth}{!}{%
{\begin{minipage}{.56\textwidth}
\setlength{\algomargin}{5mm}
\setlength{\interspacetitleboxruled}{1mm}
\begin{algorithm}[H]
	\SetKwRepeat{Do}{do}{while}
	\SetAlgoLined
\KwIn{
$I, \mu_s, \nu_k, \Bar{D}, G, \mathcal{G}_s, \mathbf{D}_{\Bar{D}}, \mathbf{G}_{s}, \mathbf{f}_{sk}, \mathbf{h}_{sk},$ $\forall k \in \mathcal{K}_s,$ $\forall s \in \mathcal{S}$\;\vspace{1mm}}
\KwOut{$\{\bm{\Theta}^*_{s1}, \cdots, \bm{\Theta}^*_{sG_s}\}$, $\forall s \in \mathcal{S}$\;}
\vspace{1mm}

\For{\normalfont{\bf each} $s \in \mathcal{S}$}{\Indmm 

Construct:
$\mathbf{\tilde{h}}_{s} = \sqrt{\mu_{s}}[\sqrt{\nu_{1}}\mathbf{\underline{h}}_{s1}^T, \cdots, \sqrt{\nu_{K_s}}\mathbf{\underline{h}}_{sK_s}^T]^T$, and

\nonl $\mathbf{\tilde{R}}_s \triangleq \sqrt{\mu_{s}}[\sqrt{\nu_1}\mathbf{\bar{R}}_{s1}^T \cdots \sqrt{\nu_{K_s}}\mathbf{\bar{R}}_{sK_s}^T]^T$,
 with: \vspace{1mm}

 \nonl \hspace{2mm}$\mathbf{\bar{R}}_{sk} = [\mathbf{R}_{sk1}\mathbf{D}_{\Bar{D}} \hspace{1mm} \cdots \hspace{1mm} \mathbf{R}_{skG}\mathbf{D}_{\Bar{D}} ]$\;

\nonl \hspace{2mm}$\mathbf{\underline{h}}_{sk} = \mathrm{vec}(\mathbf{h}_{sk}^H)$;

\nonl \hspace{2mm}$\mathbf{R}_{skg} = \mathbf{G}_{sg}^T \otimes \mathbf{f}_{skg}^H$\;

\nonl \hspace{2mm}$\mathbf{f}_{skg} = [\mathbf{f}_{sk}]_{(1 + (g-1) \Bar{D}) : g \Bar{D}}$ \;

\nonl \hspace{2mm}$\mathbf{G}_{sg} = [\mathbf{G}_s]_{(1 + (g-1) \Bar{D}) : g \Bar{D},:}$\;

\vspace{1mm}

	Initialize $\bm{\tilde{\theta}}^{(1)}_{s}$ with an arbitrary vector in the feasible set:
    $\bm{\tilde{\theta}}^{(1)}_{s} \in \{\mathbf{x} : \|\mathbf{x}\|^2 \leq \sqrt{G}, \mathbf{x} \in \mathbb{C}^{G\frac{\bar{D}(\bar{D}+1)}{2} \times 1} \}$\;\vspace{1mm}
        
		\For{$i = 1, 2, \cdots, I - 1$}{ 
		Compute the gradient of the objective function in \eqref{prob_rex_gRIS2_a_rev}: \hspace{50mm}
        $\bm{\vartheta}^{(i)}_s = 2 \mathbf{\tilde{R}}_s^H (\mathbf{\tilde{R}}_s \bm{\tilde{\theta}}^{(i)}_{s} + \mathbf{\tilde{h}}_s)$\;\vspace{1mm}
				
		Solve the direction-finding sub-problem:
  
        \nonl \hspace{1mm} $\mathbf{s}^{(i)}_s = \arg\min -(\mathbf{s}^{(i)}_s)^T \bm{\vartheta}^{(i)}_s \hspace{1mm} \text{ s.t. } \hspace{1mm} \|\mathbf{s}_s^{(i)}\|_2\leq 1$:\vspace{1mm}

        {\Indpp 
		\For{$l = 1, 2, \cdots, G \bar{D}(\bar{D}+1)/2$}{\vspace{1mm}
		$[\mathbf{s}^{(i)}_s]_{l} = \sqrt{G} \cdot \mathrm{sign}([\bm{\vartheta}^{(i)}_s]_l) \frac{|[\bm{\vartheta}^{(i)}_s]_l|}{\|\bm{\vartheta}^{(i)}_s\|_2};$
		}
  \vspace{1mm}}

        Update the step size: $\xi^{(i)} = 2/(i + 2)$\;

        Update the reflection coefficient vector: $\bm{\tilde{\theta}}^{(i+1)}_s = (1-\xi^{(i)})\bm{\tilde{\theta}}^{(i)}_s + \xi^{(i)} \mathbf{s}^{(i)}_s;$
        
        }
        \For{\normalfont{\bf each} $g \in \mathcal{G}_s$}{\Indmm
        \hspace{1mm} Compute the sub-vector for the $g$th group: $\bm{\theta}^*_{sg} = [\bm{\theta}^{(I)}]_{(1 + (g-1) \Bar{D}): g \Bar{D}}$\;
        
        \vspace{1mm}
        
        \hspace{1mm} Compute the reflection block matrix for the $g$th group: $\bm{\Theta}^*_{sg} = \mathrm{unvec}(\mathbf{D}_{\Bar{D}}\bm{\theta}^*_{sg})$\;
		}
    }
	\caption{Relaxed conditional gradient-based strategy for group-connected RIS when the direct BS-user link is available.}\label{alg_rev_2}
\end{algorithm}
\end{minipage}
}}%
\endgroup
\end{figure}

\vspace{-2mm}
{\subsection{Codebook-based practical frequency-aware configuration for group-connected RIS}\label{subsec_cb_gc}
}

Now, we can exploit the relaxed solutions $\{\bm{\Theta}^*_{sg}\}$ achieved through Algorithms \ref{alg_rev_0} or \ref{alg_rev_2}, $\forall g \in \mathcal{G}_s$ and $\forall s \in \mathcal{S}$, to determine the desired practical \ac{RIS} coefficients targeting up to $G$ distinct frequencies. This is carried out through the codebook-based strategy explained as follows.

We start by selecting the $S$ frequencies from $\mathcal{F}$ corresponding to the priority BSs, denoted as $f_s \in \mathcal{F}^{\star} \triangleq \{f_1, \cdots, f_S\} \subseteq \mathcal{F}$, such that $S \leq G \leq B$.
Then, with the help of Lemma I, we compute the relaxed transfer impedance matrix $\mathbf{Z}_{sg}^*$ associated with $\bm{\Theta}_{sg}^*$ for the $s$th BS. By exploiting the obtained matrix $\mathbf{Z}_{sg}^*$, we recall the expressions in \eqref{relax_imp_1} and \eqref{relax_imp_2}, and retrieve the associated non-redundant relaxed self and inter-element impedances, denoted by $Z^*_{sg, p}$ and $\tilde{Z}^*_{sg, pq}$, respectively. Next, for each selected frequency, we construct two codebooks $\mathcal{Z}_s = \{Z_{i}(C_{i}, f_{s})\}$ and $\mathcal{\tilde{Z}}_s = \{\tilde{Z}_{i}(\tilde{C}_{i}, f_{s})\}$ with practical self and inner impedances $Z_{i}(C_{i}, f_{s})$ and $\tilde{Z}_{i}(\tilde{C}_{i}, f_{s})$, obtained with \eqref{selfimp_eq} and \eqref{intelimp_eq}, respectively, in which $2^{B_{C}}$ capacitance values are considered in each codebook.
{With the two codebooks in hands, the desired practical impedance between elements $p$ and $q$, $\forall p < q$, with $p, q \in \{1, \cdots, \bar{D}\}$, for the $g$th RIS group assigned for the frequency $f_s$, can be achieved by
\begin{align}\label{prob_gc_cb_1}
    \hat{Z}^*_{sg, pq}(\hat{C}^*_{g, pq}, f_s) & = \underset{\forall \tilde{\zeta}_{si} \in \mathcal{\tilde{Z}}_s | i = 1, \cdots,2^{B_{C}}}{\arg \min } \hspace{2mm}
     | \tilde{Z}^*_{sg, pq} - \tilde{\zeta}_{si}
    |^2,
\end{align}
and the $p$th practical self-impedance, $\forall p \in \{1, \cdots, \bar{D}\}$, by
\begin{align}\label{prob_gc_cb_2}
    \hat{Z}^*_{sg, p}(\hat{C}^*_{g, p}, f_s) & = \underset{\forall \zeta_{si} \in \mathcal{Z}_{s} | i = 1, \cdots,2^{B_{C}}}{\arg \min } \hspace{2mm}
     | Z^*_{sg, p} - \zeta_{si}
    |^2.
\end{align}
}

\begin{figure}
\centering
\begingroup
\csname @twocolumnfalse\endcsname
\resizebox{.47\textwidth}{!}{%
\begin{minipage}{.55\textwidth}
\setlength{\algomargin}{5mm}
\setlength{\interspacetitleboxruled}{1mm}
\begin{algorithm}[H]
\SetKwRepeat{Do}{do}{while}
\SetAlgoLined

\KwIn{$\mu_s, \nu_k, D, G, \mathcal{G}_s, \mathcal{F}^\star, \mathcal{Z}_s, \mathcal{\tilde{Z}}_s, \mathbf{G}_{s}, \mathbf{f}_{sk}, \mathbf{h}_{sk},$ $\forall k \in \mathcal{K}_s,$ $\forall s \in \mathcal{S}$\;\vspace{.5mm}}
\KwOut{$\mathbf{C}_1^*, \mathbf{C}_2^*, \cdots, \mathbf{C}_G^*$\vspace{.5mm}\;}

Initialize: $\bar{D} = \frac{D}{G}$\;\vspace{1mm}

{Providing $\bar{D}$, generate the duplication matrix $\mathbf{D}_{\bar{D}}$ with Algorithm \ref{alg_dmat};}


        
        
\vspace{1mm}

{
Use $\mathbf{D}_{\bar{D}}$ and the listed input parameters to obtain the

\nonl relaxed block reflection matrices $\{\bm{\Theta}^*_{s1}, \cdots, \bm{\Theta}^*_{sG_s}\}, \forall s \in \mathcal{S}$:

\vspace{1mm}

\Indp
\uIf{direct BS-user channels are blocked}{
    With Algorithm \ref{alg_rev_0};
}\vspace{1mm}

\uIf{direct BS-user channels are available}{
    With Algorithm \ref{alg_rev_2};
}
}

\vspace{1mm}

\For{{\normalfont{\bf each} $s \in \mathcal{S}$}}{\Indmm

\For{{\normalfont{\bf each} $g \in \mathcal{G}_s$}}{\Indmm

\hspace{1mm}Obtain the corresponding relaxed impedance matrix: $\mathbf{Z}_{sg}^* = Z_0 (\mathbf{I}_{D} + \bm{\Theta}_{sg}^*) (\mathbf{I}_{D} -  \bm{\Theta}_{sg}^*)^{-1}$\;
\vspace{1mm}

{\hspace{1mm}Retrieve the non-redundant self and inter-element

\nonl \hspace{1mm}impedances, $Z^*_{sg, p}$ and $\tilde{Z}^*_{sg, pq}$, associated with $\mathbf{Z}^*_{sg}$,

\nonl \hspace{1mm}using \eqref{relax_imp_1} and \eqref{relax_imp_2};

\vspace{1mm}

\hspace{1mm}Determine the practical impedances $\hat{Z}^*_{sg, p}(\hat{C}^*_{g, p}, f_s)$ and

\nonl \hspace{1mm}$\hat{Z}^*_{sg, pq}(\hat{C}^*_{g, pq}, f_s)$, for the frequency $f_s \in \mathcal{F}^\star$, by

\nonl \hspace{1mm}exploiting the codebooks $\mathcal{Z}_s$ and $\mathcal{\tilde{Z}}_s$ in \eqref{prob_gc_cb_1} and \eqref{prob_gc_cb_2};

\vspace{1mm}

\hspace{1mm}With $\hat{Z}^*_{sg, p}(\hat{C}^*_{g, p}, f_s)$ and $\hat{Z}^*_{sg, pq}(\hat{C}^*_{g, pq}, f_s)$, finally

\nonl \hspace{1mm}obtain the matrix with the practical capacitance values

\nonl \hspace{1mm}for the $g$th group $\mathbf{\hat{C}}^*_g$.
}

}
}

\caption{Frequency-aware configuration strategy for group-connected RIS.}\label{alg2}
\end{algorithm}
\end{minipage}
}%
\endgroup
\end{figure}

{The} multi-band configuration strategy for group-connected RISs is presented in Algorithm \ref{alg_rev_2}.
{The practical impedances obtained with the codebook-based approach in \eqref{prob_gc_cb_1} and \eqref{prob_gc_cb_2} provide} a practical way to determine the required reconfigurable capacitance matrix $\mathbf{C}_g^*$ for the $g$th \ac{RIS} group, which is independent of the operating frequency. The reflection response generated by the $g$th \ac{RIS} group with $\mathbf{C}_g^*$ will be optimized for frequency $f_s$. However, such a response might lead to degraded performance for the other frequencies as a result of the frequency-dependent behavior of BD-RISs demonstrated in Section II. After determining $\mathbf{C}_g^*$, the full base-band block-diagonal scattering matrix of the group-connected \ac{RIS} for frequency $f_b \in \mathcal{F}$, can be given by $\bm{\Theta}^* (\mathbf{C}_1^*, \cdots, \mathbf{C}_G^*, f_b) = \mathrm{bdiag}(\bm{\Theta}^*_1 (\mathbf{C}_1^*, f_b), \cdots, \bm{\Theta}^*_G (\mathbf{C}_G^*, f_b))$, where the $g$th block $\bm{\Theta}_g(\mathbf{C}_g^*, f_b)$ is computed through \eqref{scatmat1}--\eqref{intelimp_eq}, $\forall g \in \{1,\cdots, G\}$. Moreover, note that if the number of \ac{RIS} groups is equal to the number of serving BSs, i.e., if $G = B$, we can set $S=B$ and become able to configure each group for each operating frequency of the network. Nevertheless, even if this is possible, one can be interested in prioritizing a subset of the frequencies (or a single frequency), where more than one group can be optimized for a common frequency. In particular, we let this flexible on purpose. In our simulation results, we shall test the performance of different group-frequency assignment criteria.

\vspace{2mm}
\section{Simulation Results}
In this section, through insightful simulation results, we investigate the impact of different operating frequencies on the performance of the proposed \ac{BD-RIS} architectures. The effectiveness of the implemented optimization strategies is also demonstrated. Moreover, we shed light on critical interference scenarios that may emerge within multi-band environments, which highlights the importance of coordination between \ac{RIS} and neighboring users and BSs.

\begin{figure}[!t]
	\centering
	\includegraphics[width=.88\linewidth]{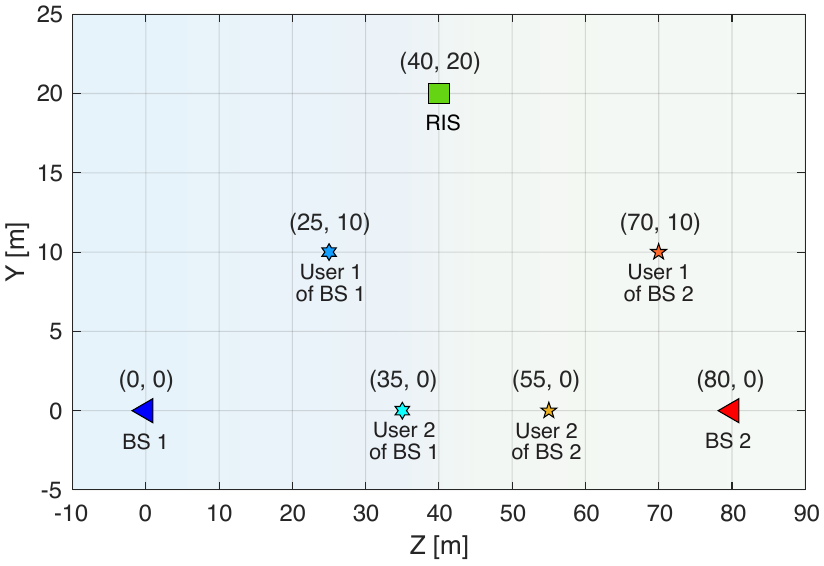}
	\caption{Standard geometrical scenario considered in the simulations, with the RIS deployed at a middle point between the two BSs. The users connected to BS $1$ are slightly closer to the RIS than those of BS $2$.\vspace{2mm}}\label{scn1}
\end{figure}

We consider a \ac{RIS}-assisted multi-band network containing $B = 2$ \acp{BS}, in which \acp{BS} $1$ and $2$ are located at the coordinates $(0, 0)$~m and $(80, 0)$~m, respectively, with each \ac{BS} employing $M = 40$ transmit antennas and serving $K_1 = K_2 = 2$ single-antenna users. Specifically, users $1$ and $2$ connected to the \ac{BS} $1$ are located at $(25,10)$~m and $(35,0)$~m, and users $1$ and $2$ connected to the \ac{BS} $2$ at $(70,10)$~m and $(55,0)$~m, respectively.
As for the \ac{RIS}, unless otherwise stated, its location is set to $(40,20)$~m. The considered simulation setup is illustrated in Fig. \ref{scn1}.
With this geometrical scenario, the path-loss coefficients for the links \ac{BS}-user, \ac{BS}-RIS, and RIS-user are computed as $(d_{bk}^{\text{\tiny \ac{BS}-U}})^{-\eta}$, $(d_b^{\text{\tiny \ac{BS}-RIS}})^{-\Tilde{\eta}}$, and $(d^{\text{\tiny RIS-U}}_{bk})^{-\Tilde{\eta}}$, respectively, where $d_{bk}^{\text{\tiny \ac{BS}-U}}$, $d_b^{\text{\tiny \ac{BS}-RIS}}$, and $d_{bk}^{\text{\tiny RIS-U}}$ are the corresponding distances, for $k\in \mathcal{K}_b = \{1, 2\}$ and $b\in \mathcal{B} = \{1, 2\}$,
{$\tilde{\eta}$ is the path-loss exponent associated with the reflected RIS links, adjusted as $2.5$, and $\eta$ is the path-loss exponent for the direct link, set to $3.5$ in the simulation examples where the direct links are available.}
Regarding the practical \ac{RIS} configuration strategies in Subsections \ref{subsec_cb_fc} and \ref{subsec_cb_gc}, the number of quantization bits is set to $B_C = 6$, in which, for generating the codebooks with self-impedances, $\mathcal{Z}$ and $\mathcal{Z}_s$, the capacitance values are uniformly spaced between $0.1$~pF and $2$~pF, while, for obtaining the codebooks with inter-element impedances, $\mathcal{\tilde{Z}}$ and $\mathcal{\tilde{Z}}_s$, the capacitances are varied between{\footnote{{Capacitance values in the femtofarad range are achievable with CMOS varactors \cite{Margalef20, Yang11}, suited for precision tuning in high-frequency applications.}}} $0.001$~pF and $.6$~pF. In addition, we adjust the remaining fixed \ac{RIS} circuit parameters in \eqref{selfimp_eq} and \eqref{intelimp_eq} as $R = \tilde{R} = 1$~$\Omega$, $L_0 = 2.5$~nH, $L = 0.7$~nH, $\tilde{L} = 0.2$~nH, $\tilde{L}_0 = 12.5$~nH, and $Z_0 = 50$. For the group-connected \ac{RIS} cases, we adjust $G = B = 2$ so that each element group can be dedicated to one distinct frequency. The weights for the users in the optimization strategies of Section \ref{RISopt} are adjusted as $\nu_1 = \nu_2 = 1/2$, whereas different values for the weights $\mu_b$ associated with the BSs are tested in our results. For fair performance comparisons, we implement the single-connected RIS circuit using the same parameters adopted in the self-impedances of the BD-RISs, and we exploit the same self-impedance codebook $\mathcal{Z}$ for performing its practical configuration.
Furthermore, we adjust the noise variance to $\sigma^2 = -40$~dBm and employ a uniform power allocation among users, such that $\alpha_{b1} = \alpha_{b2} = 1/2$, $\forall b \in \mathcal{B}$, with a total transmit power set to $P = 20$~dBm. Omitted parameters are provided in the simulation examples.

\begin{figure}[t]
	\centering
	\includegraphics[width=.95\linewidth]{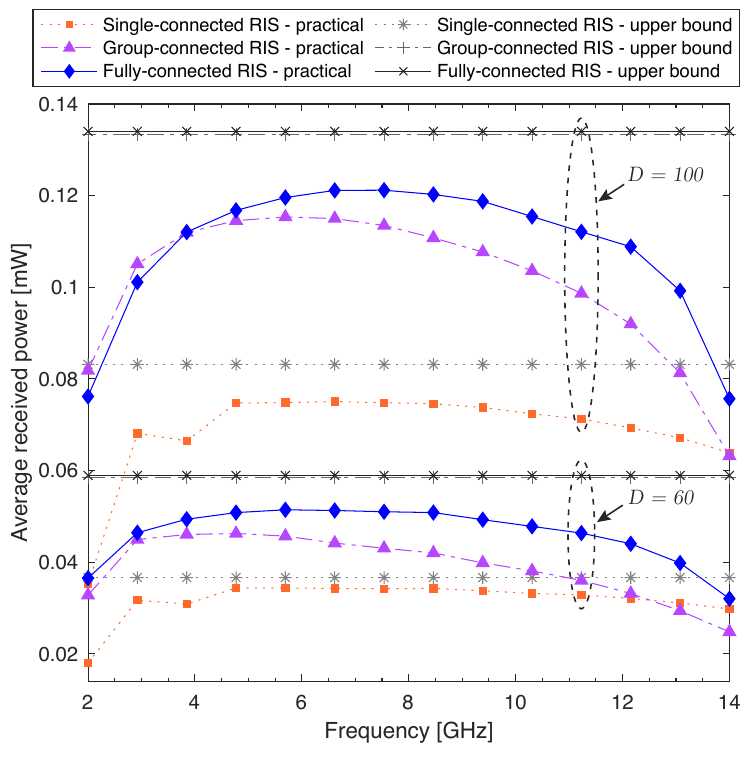}
	\caption{{Impact of the operating frequency at BS $1$ on the average received power achieved with the proposed practical configuration strategies, alongside ideal frequency-blind upper bounds from \cite{Nerini24}, for user $1$, located at $(25, 10)$~m, experiencing an obstructed direct link.\vspace{1mm}}}\label{res_rev_1}
\end{figure}

\subsection{Analysis of the frequency characteristics for different RIS architectures}

Fig. \ref{res_rev_1} presents the average received power, computed as $|\mathbf{f}_{bk}^H \bm{\Theta}\mathbf{G}_{b} \mathbf{p}_{bk} |^2 P \alpha_{bk}$, across various operating frequencies achieved with the different RIS architectures employing the proposed practical configuration strategies. For comparison, the figure also includes the corresponding upper bounds for the ideal lossless case provided in \cite{Nerini24}. The results are generated for the cases with $D=60$ and $D=100$ reflecting elements, in which, to provide clear insights, we focus on BS $1$ and study the scenario with an obstructed direct link, where only user $1$ is connected to the BS. The figure shows the preferred spectrum bands for experiencing the highest possible performances with the implemented RISs. For instance, while the fully-connected RIS yields a maximum received power of $0.12$~mW at $7.5$~GHz, at least $90\%$ of this value is still achievable from approximately $4$~GHz to $12$~GHz. In terms of relative performance, the fully-connected RIS consistently delivers the highest received power, remarkably outperforming the single-connected RIS counterpart for both $D = 60$ and $D = 100$. However, for frequencies below $4$~GHz with $D=100$, the group-connected RIS outperforms the fully-connected RIS, indicating that the circuit model for the fully-connected RIS is not optimized for lower frequencies. Aside from this range, the group-connected RIS provides an intermediate performance, as expected. Notably, the group-connected RIS can still significantly surpass the single-connected RIS across the entire considered frequency range when $D = 100$, and up to approximately $12.5$~GHz for the case with $D = 60$.

\begin{figure}[t]
	\centering
	\includegraphics[width=.95\linewidth]{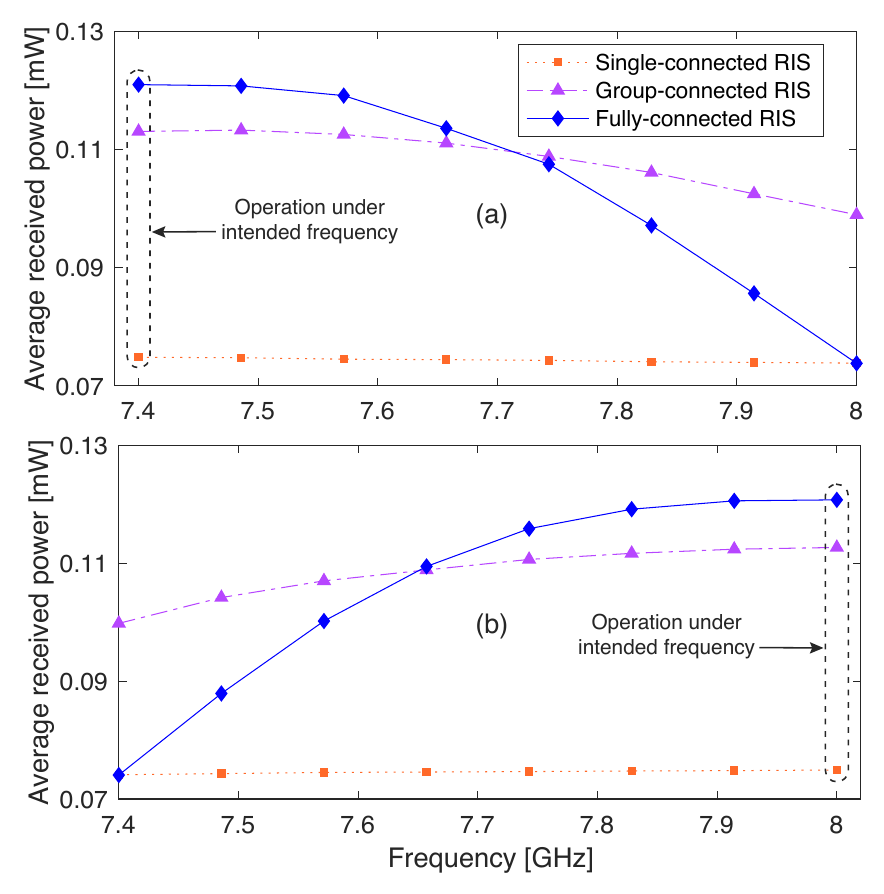}
	\caption{{Impact of operating frequency shifts at \ac{BS} $1$ on the average received power of the connected user $1$, located at $(25, 10)$~m, when the RIS is optimized only for the target frequency of (a) $7.4$~GHz, and (b) $8$~GHz, with an obstructed direct link and $D = 100$.}}\label{res1}\vspace{1mm}
\end{figure}

{Recall that while the relaxed optimization strategies for the fully-connected RIS in Section \ref{RISopt} consider the baseband channels of all frequencies, the practical codebook-based configuration is performed by targeting a single priority frequency. The deviation between the non-priority and the targeted frequency might inevitably cause performance degradation. Fig. \ref{res1} studies the impact of such frequency deviations on the average received power considering the same scenario as in the previous figure, where \ac{BS} $1$ communicates only with user $1$, assuming a blocked direct link and $D= 100$. Specifically, the relaxed optimizations in \eqref{prob_4} and \eqref{prob_rex_gRIS} are executed based on the correct baseband channels. However, the practical configuration of the \ac{RIS} is performed relying on the codebook for frequency $7.4$~GHz, in Fig. \ref{res1}(a), and $8$~GHz, in Fig. \ref{res1}(b). For comparison, the same analysis is conducted for group-connected and single-connected RISs. It can be seen in Figs. \ref{res1}(a) and \ref{res1}(b) that significant performance degradation is caused as the operating frequency deviates from the adjusted target frequency. It can also be noticed that the fully-connected \ac{RIS} is the most vulnerable architecture, experiencing the strongest decrease in received power when operating outside the target frequency. This performance behavior reveals that the proposed BD-RIS circuit models are more sensitive to frequency shifts than the conventional single-connected \ac{RIS} counterpart. Nevertheless, it is clear that, despite the highly coupled reflecting elements, BD-RISs can still effectively assist non-targeted frequencies, provided the deviation from the priority frequency is not excessively large.}

\subsection{Assessment of the proposed frequency-aware optimization strategies for BD-RISs in multi-band multi-BS environments}

\begin{figure}[t]
	\centering
	\includegraphics[width=1\linewidth]{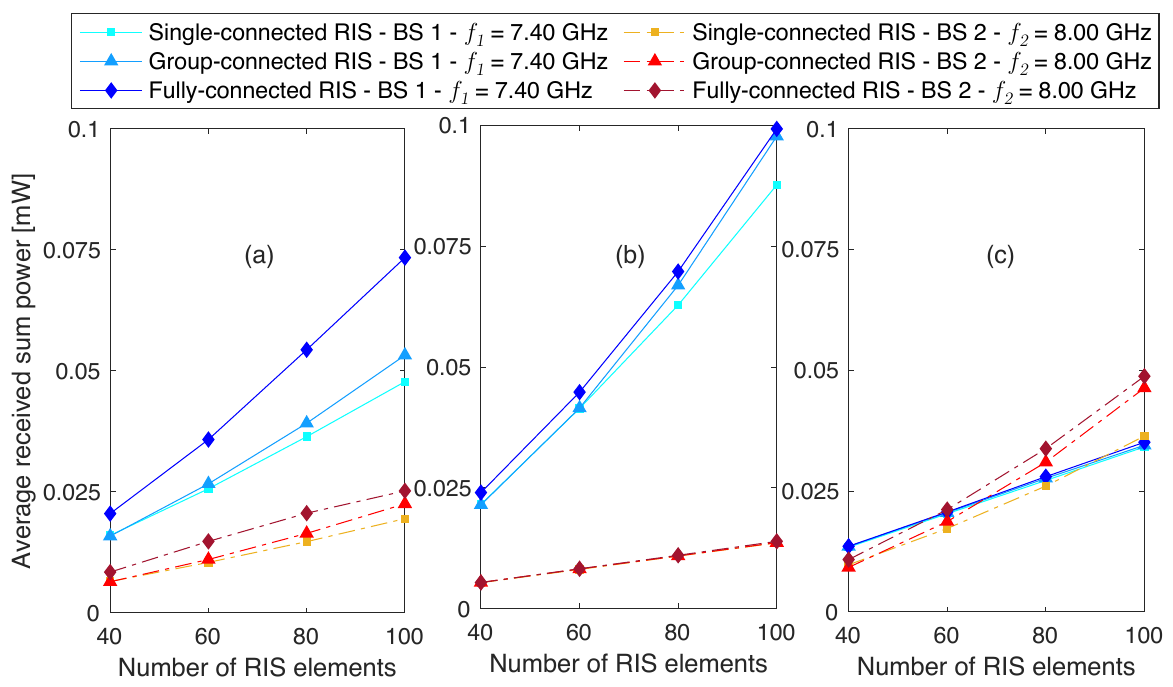}
	\caption{{Average sum of received power per \ac{BS}, with (a) $\mu_1 = 0.3$ and $\mu_2 = 0.7$; (b) $\mu_1 = 1$ and $\mu_2 = 0$; (c) $\mu_1 = 0$ and $\mu_2 = 1$, with obstructed direct links.}}\label{res2}
\end{figure}

\begin{figure}[t]
	\centering
	\includegraphics[width=1\linewidth]{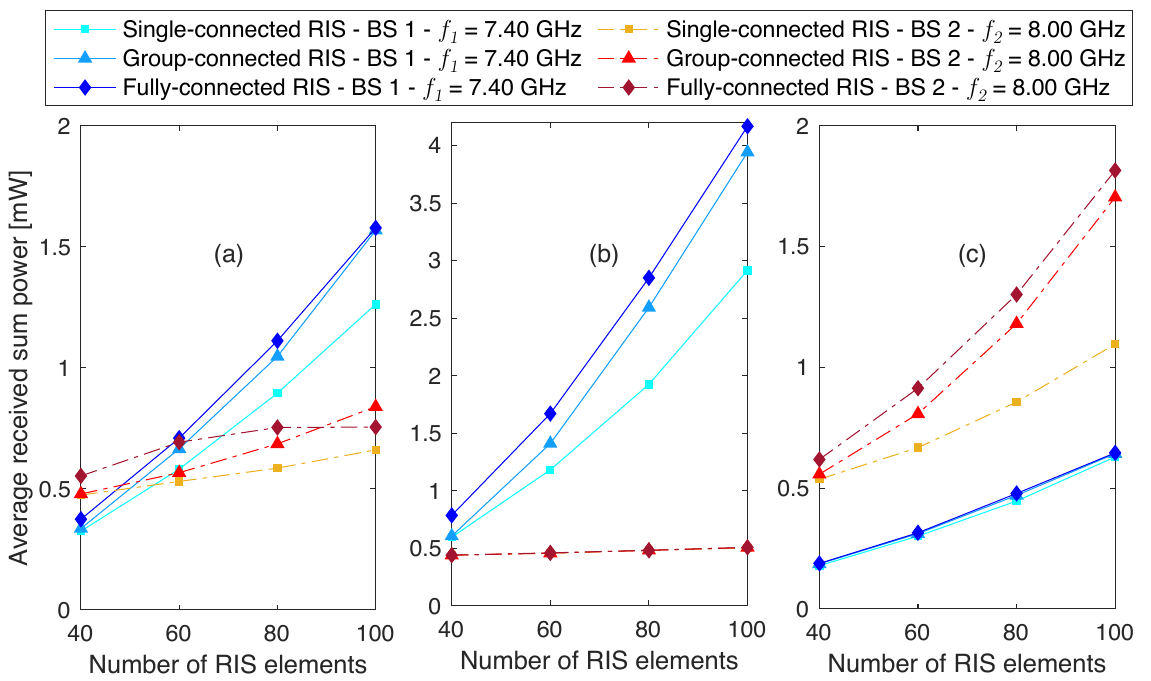}
	\caption{{Average sum of received power per \ac{BS}, with (a) $\mu_1 = 0.3$ and $\mu_2 = 0.7$; (b) $\mu_1 = 1$ and $\mu_2 = 0$; (c) $\mu_1 = 0$ and $\mu_2 = 1$,  with unobstructed direct links.}}\label{res_rev_2}
\end{figure}

\begin{figure}[t]
	\centering
	\includegraphics[width=1\linewidth]{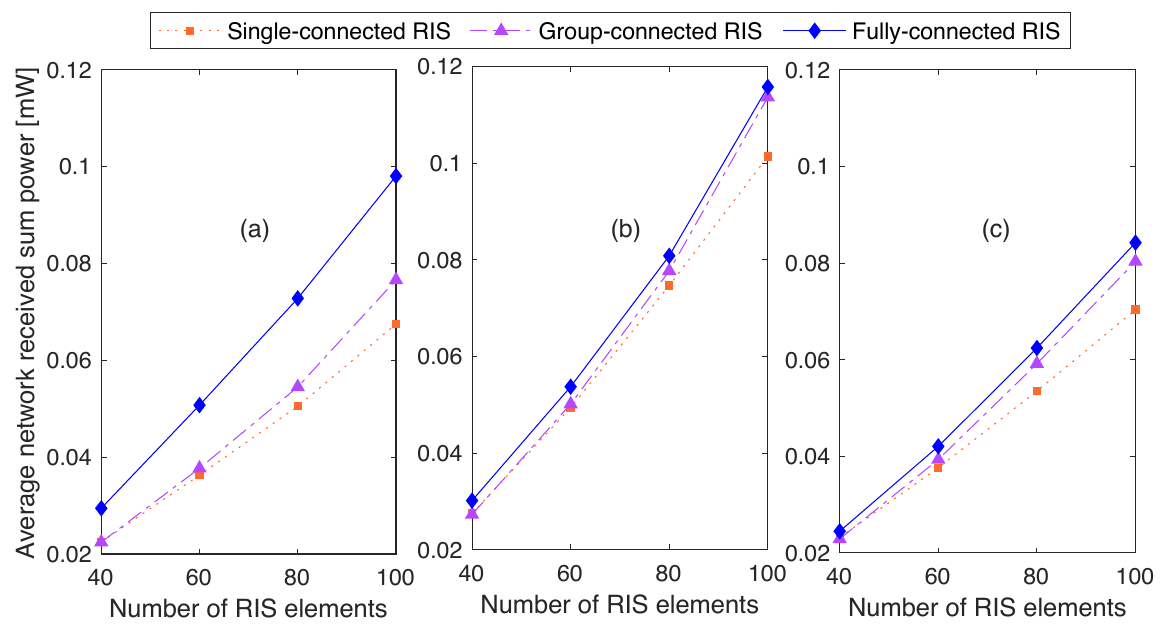}
	\caption{{Average sum of network received power, with (a) $\mu_1 = 0.3$ and $\mu_2 = 0.7$; (b) $\mu_1 = 1$ and $\mu_2 = 0$; (c) $\mu_1 = 0$ and $\mu_2 = 1$, with obstructed direct links.}}\label{res3}
\end{figure}

\begin{figure}[t]
	\centering
	\includegraphics[width=1\linewidth]{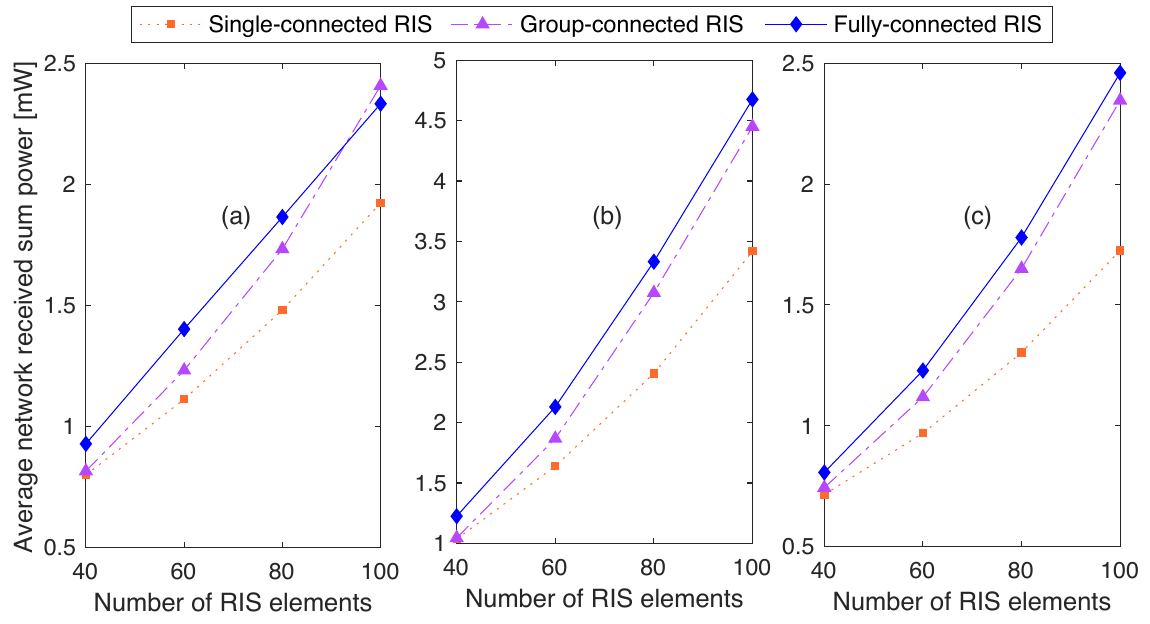}
	\caption{{Average sum of network received power, with (a) $\mu_1 = 0.3$ and $\mu_2 = 0.7$; (b) $\mu_1 = 1$ and $\mu_2 = 0$; (c) $\mu_1 = 0$ and $\mu_2 = 1$,  with unobstructed direct links.}}\label{res_rev_3}
\end{figure}

The effectiveness of the practical optimization strategies proposed in Subsections \ref{fRISopt}--\ref{subsec_cb_gc}, as well as the adaptability of these schemes for targeting multiple frequencies, is demonstrated in Figs. \ref{res2}--\ref{res_rev_3}, where the performance superiority of \acp{BD-RIS} is again verified. In Fig. \ref{res2}, the average received sum power per \ac{BS}{, for the case with blocked direct links}, i.e., $\mathrm{E}\left(\sum_{k=1}^{2}|\mathbf{f}_{bk}^H \bm{\Theta}\mathbf{G}_{b} \mathbf{p}_{bk} |^2 P \alpha_{bk}\right)$, for $b\in\{1,2\}$, is presented considering different values for the weights $\mu_b$. Specifically, in Fig. \ref{res2}(a), motivated by the fact that the users connected to \ac{BS} $2$ are further away from the \ac{RIS} than those of \ac{BS} $1$, we set a higher weight for \ac{BS} $2$, i.e., $\mu_2 = 0.7$, so that a more balanced performance can be achieved. Following the strategy of \cite{Cai21}, we configure the single connected \ac{RIS} by dedicating half of the elements to frequency $f_1 = 7.4$~GHz and the other half to frequency $f_2 = 8$~GHz. As for the group-connected \ac{RIS} case, we assign each \ac{RIS} group to one distinct frequency. On the other hand, following the strategy proposed in Subsection \ref{fRISopt}, the relaxed coefficients for the fully-connected \ac{RIS} are optimized based on the baseband channels of all users of the two BSs. However, because a practical fully-connected \ac{RIS} can be dedicated to up to one target frequency, its practical coefficients are determined based on the codebook for the frequency $f_1 = 7.4$~GHz. As can be seen in Fig. \ref{res2}(a), despite this constraint, the fully-connected \ac{RIS} can still achieve the highest performance under both considered operating frequencies. This result indicates that, at the cost of higher hardware and optimization complexity, a fully-connected \ac{RIS} is the most robust among the considered architectures to multi-user multi-band scenarios undergoing blocked direct links, despite the fact that its practical configuration must be targeted at a single frequency. The group-connected \ac{RIS}, on the other hand, provides an overall reduced complexity but can still outperform the single-connected \ac{RIS} counterpart, even though lower gains are achievable. This result is aligned with the findings of \cite{Shen22}, confirming that group-connected RISs offer a trade-off between complexity and performance.

In Fig. \ref{res2}(b), we set $\mu_1 = 1$ and $\mu_2 = 0$ and optimize the \ac{RIS} coefficients exclusively for \ac{BS} $1$ operating with $f_1 = 7.4$~GHz. On the one hand, we can see that the received sum power of users of \ac{BS} $1$ is considerably improved, especially for the \ac{BD-RIS} cases. On the other hand, low power levels are delivered to the users of \ac{BS} $2$, with all \ac{RIS} architectures achieving nearly the same performance. This behavior results from the fact that the \ac{RIS} operates as it is employing random coefficients for \ac{BS} $2$ since the channels or its connected users are not being taken into account in the optimization, i.e., $\mu_2 = 0$. We study the opposite scenario in Fig. \ref{res2}(c), in which, by considering $\mu_1 = 0$ and $\mu_2 = 1$, we optimize the RISs focused on the \ac{BS} $2$ only, the \ac{BS} with the weakest users which are being served under $f_2 = 8$~GHz. In this case, both fully-connected and group-connected RISs can deliver significant received power gains to the users of \ac{BS} $2$ when compared to the corresponding curves in Figs. \ref{res2}(a) and \ref{res2}(b). However, the achieved improvements are not as high as those for \ac{BS} $1$ in Fig. \ref{res2}(b) due to the higher path losses in the RIS reflected links for the users of \ac{BS} $2$. Moreover, a significant decline in the sum received power for \ac{BS} $1$ can be observed for all architectures as a result of the random coefficients employed for the users of \ac{BS} $1$ when $\mu_1 = 0$.

{The scenario with available direct links is investigated in Fig. \ref{res_rev_2}, where the average received sum power per \ac{BS} is calculated as $\mathrm{E}\left(\sum_{k=1}^{2}|(\mathbf{f}_{bk}^H \bm{\Theta}\mathbf{G}_{b} + \mathbf{h}_{bk}^H) \mathbf{p}_{bk} |^2 P \alpha_{bk}\right)$. Although BD-RISs continue to excel also in this scenario, these results bring new insights. The first difference we can notice is that the received power levels are significantly higher than those observed in Fig. \ref{res2}. Moreover, in Fig. \ref{res_rev_2}(a), the fully-connected RIS exhibits a performance behavior that is not observed with blocked direct links. As can be seen, even though the fully-connected RIS offers high performance for the priority BS $1$ across all numbers of reflecting elements, the received power at the users of BS $2$ with the fully-connected RIS saturates, with the group-connected RIS achieving the best performance when the number of reflecting elements reaches $D = 100$. Such a behavior is a consequence of the fact that fully-connected RISs are more sensitive to frequency shifts as revealed in Fig. \ref{res1}, a characteristic that is boosted with the availability of the direct link. In Figs. \ref{res_rev_2}(b) and \ref{res_rev_2}(c), we study the tradeoffs of dedicating the RISs to individual BSs, as performed for the scenario with blocked direct links. As before, we can also see that when $\mu_1 = 1$ and $\mu_2 = 0$, the sum power for BS $1$ reaches the highest values due to the closer proximity of the associated users to the RIS. However, in contrast to Fig. \ref{res2}(c), we can observe more significant improvements for BS $2$ in Fig. \ref{res_rev_2}(c), with $\mu_2 = 1$ and $\mu_1 = 0$, thanks to the contribution of the direct channel link.}

{Lastly, we investigate the impact of the choices of $\mu_b$ on the overall performance of the network. To this end, we plot the network average received sum power computed as $\mathrm{E}\left(\sum_{b=1}^{2}\sum_{k=1}^{2}|\mathbf{f}_{bk}^H \bm{\Theta}\mathbf{G}_{b} \mathbf{p}_{bk} |^2 P \alpha_{bk}\right)$ for scenarios with blocked direct links, and $\mathrm{E}\left(\sum_{b=1}^{2}\sum_{k=1}^{2}|(\mathbf{f}_{bk}^H \bm{\Theta}\mathbf{G}_{b} + \mathbf{h}^H_{bk})\mathbf{p}_{bk} |^2 P \alpha_{bk}\right)$ for scenarios with available direct links, in Figs. \ref{res3} and \ref{res_rev_3}, respectively. In Fig. \ref{res3}(a), employing balanced weights,} each \ac{RIS} architecture is configured to serve the two BSs simultaneously. As a result, an intermediate network performance is achieved in comparison to the subsequent subfigures. Fig. \ref{res3}(b), in turn, provides the highest received sum power as a result of the fact that the \ac{RIS} is dedicated to \ac{BS} $1$ which counts with the closest users. In the other extreme, Fig. \ref{res3}(c) brings the lowest network sum powers as only \ac{BS} $2$ is being considered in the \ac{RIS} optimizations.
{Similar trends can be observed in Fig. \ref{res_rev_3}, where direct links are available. However, when employing balanced weights in Fig. \ref{res_rev_3}(a), the group-connected RIS outperforms the fully-connected RIS counterpart at the network level when $D=100$, reflecting the behavior described in Fig. \ref{res_rev_2}(a). Nevertheless, the performance achieved by the two BD-RIS architectures is still remarkably dominant, outperforming the single-connected RIS in all tested cases across Figs. \ref{res_rev_3}(a)--(c).} 
These results show that the overall network performance is highly influenced by the value of $\mu_b$, which should be carefully selected based on the requirements and objectives of the system. For instance, if the goal is to maximize network received sum power, the highest weight must be given to the frequency in which the strongest users are operating. If the goal is to assist users connected to different BSs under distinct frequencies instead, balanced weights should be preferred. Furthermore, it is clear that the availability of direct links plays a significant role and strongly influences the network performance.

\begin{figure}[!t]
	\centering
	\includegraphics[width=.88\linewidth]{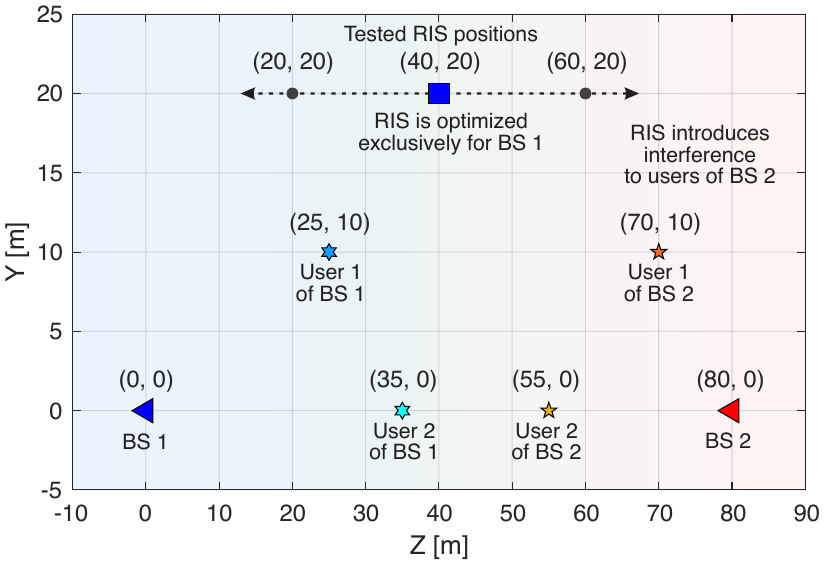}
	\caption{Illustration of the tested interference scenario when the RIS is not synchronized with the BS $2$. The RIS is placed at three different coordinates.}\label{scn2}
\end{figure}

\begin{figure}[!t]
	\centering
	\includegraphics[width=.95\linewidth]{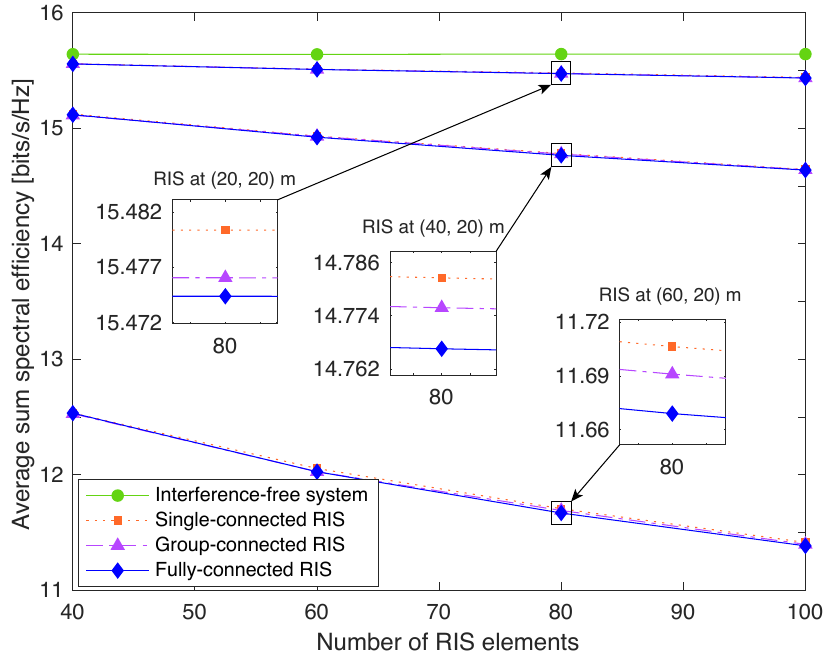}
	\caption{{Average sum spectral efficiency of users connected to \ac{BS} $2$ located at $(0, 80)$~m under frequency $f_2 = 8.4$~GHz when the \ac{RIS} is optimized targeting \ac{BS} $1$ under frequency $f_1 = 7.4$~GHz, with $\mu_1 = 1$ and $\mu_2 = 0$.}}\label{res4}
\end{figure}

\subsection{Analysis of the unplanned deployment of RISs in multi-band multi-BS environments}

In the previous results, we have seen that if an \ac{RIS} is dedicated entirely to one of the BSs (e.g., $\mu_1 = 1$ and $\mu_2 = 0$, or vice-versa), the \ac{RIS} coefficients behave as random for the other \ac{BS} employing a different frequency. Nonetheless, it can be noticed that, even though the received power for the non-intended \ac{BS} is degraded, it is still non-zero. This provides a strong indication that if RISs are not properly deployed or well coordinated with neighboring BSs, non-negligible interference can be introduced and degrade multiple access performance. This critical issue is confirmed in Fig. \ref{res4}, which presents the average sum spectral efficiency for \ac{BS} $2$, with frequency $f_2 = 8.4$~GHz, under the impact of non-intended reflections induced by an \ac{RIS} that is optimized exclusively for \ac{BS} $1$ under the target frequency $f_1 = 7.4$~GHz, i.e., $\mu_1 = 1$ and $\mu_2 = 0$. 
This investigated interference scenario is illustrated in Fig. \ref{scn2}.
For this example, specifically, we assume that the users connected to \ac{BS} $2$ experience a good signal reception in the direct \ac{BS}-U link and that \ac{BS} $2$ is not aware of the deployment of the \ac{RIS}. The channel estimate acquired by \ac{BS} $2$ becomes, consequently, outdated due to the unsynchronized operation of the \ac{RIS}. As a result, the precoders of \ac{BS} $2$ cannot cancel perfectly inter-user interference. Different \ac{RIS} locations are tested to study the impact of such harmful reflections. As can be seen, even at the coordinate $(20,20)$~m, where the \ac{RIS} is relatively far from the users of \ac{BS} $2$, a noticeable degradation in the sum spectral efficiency curves can be already observed under all three \ac{RIS} architectures. This performance degradation significantly intensifies as the \ac{RIS} comes closer to the \ac{BS} $2$ and associated users, resulting in a substantial gap of nearly $4$~bits/s/Hz between the sum spectral efficiency achieved by the interference-free system and the case with the \ac{RIS} comprising $D = 80$ elements located at $(60,20)$~m. Moreover, despite all architectures exhibiting similar performance degradation, when the curves are zoomed in, we can observe that the fully-connected \ac{RIS}, owing to its superior reflection efficiency, causes slightly stronger degradation, while the single-connected \ac{RIS} exhibits the mildest impact.
These findings highlight the necessity of coordination and synchronization in multi-band multi-BS networks to avoid undesired performance degradation,  regardless of the employed architecture.

\section{Conclusions}
This paper has addressed an important gap in the existing literature on \acp{BD-RIS} by carrying out a novel investigation into their frequency-dependent behavior. We proposed a new frequency-dependent reflection model applicable for both fully-connected and group-connected \acp{RIS}, based on which an efficient and practical framework for configuring these promising devices in multi-band multi-\ac{BS} \ac{MIMO} environments was developed.
Specifically, by relying on a codebook-based approach, integrated with a low-complexity matrix theory-based solution, we implemented flexible multi-objective optimization schemes capable of maximizing the received power at multiple users served under different frequencies. The effectiveness of our strategies was validated through comprehensive simulations across various scenarios. 
Our results have not only revealed the frequency-dependent performance of different \ac{RIS} architectures but also demonstrated the superiority of \acp{BD-RIS} over conventional single-connected \ac{RIS} counterparts. Furthermore, our findings stress the critical importance of regulating the placement and coordinating the operation of RISs and BSs in multi-band scenarios in order to counter harmful interference. Such measures are crucial to avoid deteriorated performance, emphasizing the need for meticulous planning and optimization for the effective deployment of \ac{RIS} technology in future 6G.


\appendices

\section{Proof of Proposition I}\label{ap1}
\renewcommand{\theequation}{A-\arabic{equation}}
\setcounter{equation}{0}

Note that the product $\mathbf{D}_D \bm{\theta} \in \mathbb{C}^{D^2 \times 1}$ results an augmented vector in which, out of $\frac{D(D+1)}{2}$ entries of $\bm{\theta}$, $\frac{D(D-1)}{2}$ elements are duplicated (the off-diagonal entries of $\bm{\Theta}$), and $D$ elements are kept non-redundant (the main diagonal entries of $\bm{\Theta}$). This implies that for any $\bm{\theta}$, the $L_2$ norm of $\mathbf{D}_D\bm{\theta}$ will always be less than twice the norm of $\bm{\theta}$, as a result of the fact that not all of its elements are duplicated by $\mathbf{D}_D$. Mathematically, the following inequality holds
\begin{align}\label{prop_lii}
    \|\mathbf{D}_D \bm{\theta} \|_2 &= \sqrt{2 \sum_{\forall i > j} |[\bm{\Theta}]_{ij}|^2 + \sum_{\forall i = j} |[\bm{\Theta}]_{ij}|^2} \nonumber\\
    & < \sqrt{2 \sum_{\forall i \geq j} |[\bm{\Theta}]_{ij}|^2} = \sqrt{2} \| \bm{\theta} \|_2.
\end{align}
With property \eqref{prop_lii} in hand, and by assuming that $D \geq 2$, it is guaranteed that
\begin{align}\label{prop2_lii}
\frac{1}{\sqrt{D}}\|\mathbf{D}_D\bm{\theta} \|_2 < \frac{{\sqrt{2}}}{\sqrt{D}} \| \bm{\theta} \|_2 \leq \| \bm{\theta} \|_2.    
\end{align}

The inequality in \eqref{prop2_lii} ensures that $\| \bm{\theta} \|_2 > \frac{1}{\sqrt{D}}\|\mathbf{D}_D\bm{\theta} \|_2$. As a result, it follows that by constraining $\| \bm{\theta} \|_2 \leq 1$, the inequality $\frac{1}{\sqrt{D}}\|\mathbf{D}_D \bm{\theta}\|_2 < 1$ holds with probability one, which completes the proof.  \hfill \qedsymbol

{\section{Proof of Lemma I}\label{ap2}}
\renewcommand{\theequation}{B-\arabic{equation}}
\setcounter{equation}{0}

First, let the expression in \eqref{scatmat1} be rewritten as follows:
\begin{align}
    (\mathbf{Z} + Z_0\mathbf{I}_{D})\bm{\Theta}  &= (\mathbf{Z} + Z_0\mathbf{I}_{D})(\mathbf{Z} + Z_0\mathbf{I}_{D})^{-1} (\mathbf{Z} - Z_0\mathbf{I}_{D}) \nonumber\\
    &=  (\mathbf{Z} - Z_0\mathbf{I}_{D}).
\end{align}
Then, through simple algebraic manipulation, the following can be achieved
\begin{align}\label{impedmat12}
    \mathbf{Z}\bm{\Theta} - \mathbf{Z} \mathbf{I}_{D} &= 
 - Z_0 \bm{\Theta} - Z_0\mathbf{I}_{D}. 
\end{align}
From \eqref{impedmat12}, we can readily derive the desired expression for $\mathbf{Z}$, as shown in \eqref{impedmat1}, which completes the proof. \hfill \qedsymbol \vspace{2mm}


\ifCLASSOPTIONcaptionsoff
\newpage
\fi

\bibliographystyle{IEEEtran}
\bibliography{TWC_Preprint_V2}

\end{document}